\documentclass[preprint,authoryear]{elsarticle}

\usepackage{adjustbox}
\usepackage{amsfonts}
\usepackage{amsthm}
\usepackage{amsmath}
\usepackage{bm}
\usepackage{tabularx}

\usepackage{natbib}

\usepackage{algorithm}
\usepackage{algorithmic}

\usepackage{graphicx}
\usepackage{makecell,rotating}
\usepackage[labelsep=quad,indention=10pt]{caption}
\usepackage[list=true]{subcaption}
\usepackage{hyperref}

\DeclareCaptionSubType*[arabic]{table}
\DeclareCaptionSubType*[arabic]{figure}
\usepackage{chngcntr}
\usepackage{color}

\usepackage{bigints}

\makeatletter
\newcommand*{\rom}[1]{\expandafter\@slowromancap\romannumeral #1@}
\makeatother
\usepackage{amsfonts}
\usepackage{amsthm,amssymb,amsmath}

\theoremstyle{plain}

\theoremstyle{definition}

\theoremstyle{remark}

\newcommand{\N}{\ensuremath{\mathbb{N}}}   
\newcommand{\R}{\ensuremath{\mathbb{R}}}   

\def \by {{\bf y}}
\def \Y {\mathcal{Y}} 
\def \R {\mathbb{R}}
\def \bw {\mathbf{w}}
\def \E {\mathbb{E}}
\def \W {\mathcal{W}}
\def \N {\mathcal{N}}
\def \diag {\textup{diag}}
\def \bx {\mathbf{x}}
\def \bvec {\textup{vec}}
\def \bg {\mathbf{g}}
\def \divg {\textup{div}}
\def \Divg {\textup{Div}}
\def \bv {\mathbf{v}}
\def \btheta {{\bm\theta}}
\def \balpha {{\bm\alpha}}
\def \bepsilon {{\bm\epsilon}}

\usepackage{fancyhdr}
\begin{document}

\title{Bayesian Forecast Combination with Predictive Priors via Particle Filtering}

\author[1]{Xiaorui Luo}
\affiliation[1]{organization={School of Mathematical Sciences, Beihang University (Shahe Campus)},
            addressline={Changping district}, 
            city={Beijing},
            postcode={102206}, 
            state={P. R.},
            country={China}}
\ead{xruiluo@buaa.edu.cn}

\author[2]{Yanfei Kang}
\affiliation[2]{organization={School of Economics and Management, Beihang University},
            addressline={Haidian district}, 
            city={Beijing},
            postcode={100191}, 
            state={P. R.},
            country={China}}
\ead{yanfeikang@buaa.edu.cn}

\author[1,3]{Xue Luo\corref{cor1}}
\affiliation[3]{organization={Key Laboratory of Mathematics, Informatics and Behavioral Semantics (LMIB)},
            addressline={Haidian district}, 
            city={Beijing},
            postcode={100191}, 
            state={P. R.},
            country={China}}
\ead{xluo@buaa.edu.cn}

\cortext[cor1]{Corresponding author.}

\begin{abstract}
We propose a Bayesian forecast combination framework that, for the first time, embeds forward-looking signals—formulated as predictive priors—directly into the time-varying weight-updating process. This approach enables weights to adapt using both historical forecast performance and anticipated future model behavior. We implement the framework with model diversity as the forward-looking signal, yielding the diversity-driven time-varying weights (DTVW) method. Compared with the standard time-varying weights (TVW) approach, DTVW embeds diversity-driven predictive priors that penalize redundancy and encourage informative contributions across constituent models. Simulation experiments, covering both a simple complete model set and a complex misspecified environment, show that DTVW improves forecast accuracy by dynamically focusing on well-performing models. Empirical applications to multi-step-ahead oil price forecasts and bivariate forecasts of U.S. inflation and GDP growth confirm its superiority over benchmarks including Equal weighting, Bayesian Model Averaging, and standard TVW. Beyond accuracy gains, diversity-based predictive priors provide diagnostic insights into model incompleteness and forecast uncertainty, making DTVW both more adaptive and more informative than existing Bayesian combination methods.
\end{abstract}

\maketitle

\section{Introduction}

Forecast combinations derived from multiple models or methods are often more robust than relying on a single ``best" forecast, as they integrate diverse sources of information and help mitigate model-specific weaknesses. The seminal work by \cite{BG:69} laid the foundation for forecast combination. Subsequently, \cite{C:89} provided systematic reviews of its theoretical underpinnings and empirical findings. More recent advancements span from simple averaging to more sophisticated model-based frameworks~\citep{WHLK:23}, all driven by the same principle: integrating diverse perspectives improves accuracy and reliability.

Among the various approaches to forecast combination, Bayesian methods offer a principled probabilistic framework. Bayesian Model Averaging (BMA), as outlined in \cite{HMRV:99}, provides a formal probabilistic framework for combining models based on their posterior model probabilities. Compared to point forecasts, BMA characterizes uncertainty probabilistically, offers richer insights into forecast risks, and facilitates more optimal decision-making under uncertainty~\citep{RGBP:05,HM:07,GR:13}. More recently, Bayesian Predictive Synthesis (BPS), developed by \cite{MW:19} and further extended by \cite{MANW:20}, builds on earlier work in Bayesian agent or expert opinion analysis~\citep{W:84,GS:85,WC:92}, integrating data-driven updating mechanisms to account for forecast dependencies and model uncertainty. In contrast to BMA and BPS, another type of methods stems from a distinct Bayesian perspective on forecast combination. It is grounded in the idea that combination weights can be treated as latent variables evolving over time within a state-space framework. This line of research traces back to \cite{TV:02} and was further advanced in Dynamic Model Averaging (DMA) by \cite{RAKME:10}, which employs forgetting factors and recursive Kalman filtering for efficient online updating. More recently, \cite{BCMP:24} extended DMA to Bayesian Dynamic Quantile Model Averaging, leveraging sequential Monte Carlo for time-varying quantile regression. Our method follows this line and is most closely related to the work of \cite{B:13} and \cite{CRV:15}, who assumed that combination weights obey a latent random process within a general distributional Bayesian state-space framework. 

Bayesian analysts emphasize the value of incorporating predictive information into prior specification. While traditional priors are often chosen for mathematical convenience or derived from subjective beliefs, predictive priors explicitly embed informed judgments about future or hypothetical data outcomes \citep{GVC:20}. This perspective aligns with \citet{GSB:17}, who advocate grounding priors in observable data patterns. For example, \citet{V:25} note that experts can often articulate beliefs about predictive quantities—such as specific quantiles or the probability of extreme events—which can be translated into priors via the prior predictive distribution. By calibrating priors to ensure coherence with plausible predictive scenarios, predictive priors can enhance both interpretability and performance, particularly in applied settings where prior information is intrinsically data-driven or simulation-based \citep{GSVBG:19,HABK:20}.

However, this forward-looking principle in Bayesian prior design has seen little application in forecast combination, where weight estimation remains almost entirely anchored to historical performance.  For instance, \cite{B:13} used an exponentially weighted moving average of past forecast errors as a learning mechanism. Building on such work, \cite{DMHBS:16} expanded the filtration generated by historical data and allowed hyper-parameters to be updated over time together with weights, but still without incorporating future information. Although \cite{CA:24} assigned weights by anticipating future forecast errors and minimizing the associated expected loss, their method is non-Bayesian.

In this paper, we introduce a forward-looking feedback framework for Bayesian forecast combination, in which the prior process for weights incorporates both historical observations and forward-looking signals on model performance. This design enables the model to anticipate structural changes and adjust weights accordingly, improving both accuracy and stability. By reducing over-reliance on historical data—a common limitation of conventional approaches—this framework promotes a balanced weighting scheme that blends past performance with prospective indicators. As a result, this combination forecast becomes more flexible and robust in the face of evolving dynamics.

To operationalize this framework, we employ model diversity as a forward-looking signal. Diversity penalizes redundancy and captures heterogeneity in models’ expected predictive behavior. While the framework can accommodate alternative forward-looking measures tailored to specific applications, diversity is a natural and well-studied choice in the forecast combination literature \citep{BG:69,BD:95,T:19,AMRV:19}.
\citet{KW:22} provided a quantitative definition of diversity and demonstrated its importance, and \citet{G:23}  observed that prioritizing diversity over historical accuracy often enhances ensemble performance. However, an exclusive focus on diversity may lead to overvaluing models with persistently poor forecasts simply because they differ from their better-performing counterparts. Conversely, a sole emphasis on past forecast accuracy risks overlooking models with subpar individual performance that nonetheless offer valuable, unique insights. Striking this balance is therefore essential to achieve robust and accurate forecast combinations.

We propose a diversity-driven time-varying weight (DTVW) approach to forecast combination that implements a forward-looking feedback mechanism by incorporating model diversity as future-oriented information. Our DTVW builds upon the time-varying weight (TVW) method proposed by \cite{B:13}. The weights in DTVW are governed by a latent process and formulated as a regression incorporating both historical forecast information and forward-looking signals, with all regression coefficients learned sequentially from the data. This approach encourages the integration of more diverse information sources into the forecast combination by assigning higher weights to models that offer greater forecast diversity. In both simulation and empirical applications, the estimated time-varying parameters preceding the diversity term are consistently positive-consistent with the intuition that greater diversity (reflected in lower correlation among models) is beneficial. This mechanism effectively penalizes redundant information and facilitates the integration of complementary signals, thus enhancing the robustness and accuracy of the combined forecast.

Two simulation experiments are performed to assess the effectiveness of DTVW. The first experiment assumes a simple complete model set, replicating the setup of \cite{B:13} with three AR models that have distinct unconditional means. Our DTVW enhances forecast accuracy while retaining the capability to identify the true model. Additionally, with appropriate initialization of parameters in the latent process, the DTVW degenerates to the TVW when the diversity term is artificially eliminated.
The second experiment considers an incomplete and nonlinear data-generating process, where the true model is not included in the candidate set. Here, our DTVW again outperforms the TVW in terms of accuracy and effectively identifies the top two well-performing models by assigning them higher weights, while TVW mistakes one candidate as the true model.

We further apply DTVW to two empirical settings with contrasting characteristics. The first dataset from \cite{KJH:23} consists of monthly Brent crude oil prices, which are characterized by high volatility, structural breaks, and strong non-stationarity. Across all forecast horizons, our DTVW consistently outperforms competing approaches (e.g., TVW and BMA), showcasing strong performance in challenging and unstable environments.
The second dataset comprises quarterly U.S. macroeconomic variables-specifically, inflation (measured by the PCE deflator) and real GDP growth-which exhibit smoother and more persistent patterns. Using the same data as in \cite{B:13}, we conduct bi-variate joint density forecasts to evaluate our DTVW, which again yields consistently better results. By incorporating model diversity as a predictive prior, the combination weights adapt dynamically: assigning greater weight to better-performing models while down-weighting those with poor predictive accuracy. This underscores the method’s flexibility and reliability in both volatile energy markets and stable macroeconomic environments.

Our contributions are twofold. First, we extend the Bayesian forecast combination framework by embedding predictive priors directly into the latent weight process through a forward-looking signal. This design enables the model to anticipate structural changes, adapt more rapidly to evolving conditions, and reduce forecast uncertainty. The framework is highly flexible: the forward-looking signal can be constructed in various ways to suit different application contexts, allowing broad adaptability and extensibility.
Second, we operationalize this framework using a model diversity metric as the forward-looking signal, leading to the proposed DTVW method. By penalizing high inter-model correlation and promoting information complementarity, DTVW adaptively allocates weights to maximize the value of unique predictive contributions. To our knowledge, this is the first approach to incorporate forward-looking signals within a Bayesian weight-updating process, enabling forecast combinations to look ahead and prepare for future conditions rather than relying solely on extrapolation from historical data.

The paper is structured as follows. Section \ref{sec 2} introduces the Bayesian framework for forecast combination and defines the scaled model diversity. Section \ref{sec-3} elaborates on the DTVW, detailing how diversity is incorporated as a forward-looking element within the aforementioned framework, and provides a detailed account of its implementation via particle filtering. Section \ref{sec-4} presents simulation experiments, including accuracy comparisons and the ability to identify well-performing models. Section \ref{sec-5} offers empirical applications using economic data, focusing on GDP, inflation, and oil prices. Section \ref{sec-6} concludes and puts forward suggestions for future research.

\section{Preliminary}\label{sec 2}
\setcounter{equation}{0}

In this section, we lay the foundational groundwork by formalizing the Bayesian framework for forecast combination and introducing the concept of model diversity. The Bayesian framework for forecast combination treats weights as random variables with priors, integrating prior knowledge and data to synthesize model outputs. Weights reflect model credibility and update dynamically, following Bayesian sequential learning. Model diversity, quantifying forecast dissimilarity, captures non-redundant information to complement accuracy. Operationalized via prediction discrepancies, it embeds into the framework to balance history and forward-looking complementarity, enabling our DTVW approach in Section \ref{sec-3.2}.

\subsection{The Bayesian framework of the combination forecast}\label{sec-2.1}

Assume that $\by_t=(y_{t}^{1},y_{t}^{2},\cdots,y_{t}^{L})\in\Y\subset\R^L$ is the observable variable at time $t$. We denote $\by_{1:t-1}=\{\by_1,\cdots,\by_{t-1}\}$ as a sequence of the $L$-vector of the observable from $1,\cdots,t-1$. The ultimate interest in the density forecast is to achieve/approximate the predictive distribution of the observable based on the observables' history, denoted as $p(\by_t|\by_{1:t-1})$. In the realm of the combination forecast, one further assumes that there are $K$ predictors of $\by_t$, denoted as $\tilde\by_{k,t}=(\tilde y^1_{k,t},\cdots,\tilde y^L_{k,t})\in\Y\subset\R^L$, $k=1,\cdots,K$. Presumably, the predictive density for each predictor $\tilde \by_{k,t}$ is $p(\tilde \by_{k,t}|\by_{1:t-1})$, $k=1,\cdots,K$. For short of notations, we denote $\tilde\by_t=(\tilde\by_{1,t}',\cdots,\tilde\by_{K,t}')\in\R^{L\times K}$, with $'$ representing the transpose. Each column is a predictor of $\by_t$. Let $\tilde \by_{1:t}=\{\tilde\by_1,\cdots,\tilde \by_t\}$ be a sequence of $L\times K$ matrices.


Naturally, the relationship between the density of $\by_t$ conditionally on $\by_{1:t-1}$ and the collection of the predictive densities from the $K$ different models is given by
\begin{equation}\label{eqn-2.1}
    p(\by_t|\by_{1:t-1})=\int_{\Y^{K\times t}}p(\by_t|\tilde\by_{1:t},\by_{1:t-1})p(\tilde \by_{1:t}|\by_{1:t-1})d\tilde\by_{1:t},
\end{equation}
where the joint predictive density of the predictors at time $t$ is assumed to only depend on the observables up to time $t-1$, i.e.,
\begin{equation}\label{eqn-2.2}
    p(\tilde{\mathbf{y}}_{1:t} | \by_{1:t-1}) = \prod_{s=1}^{t} p(\tilde{\mathbf{y}}_s | \by_{1:s-1}).
\end{equation}
For the sake of simplicity, we only state the one-step ahead forecast, but it can be easily extended to the multiple-step ahead scenario. 

Within this combination forecast framework, the weight specification of the $K$ predictors $\tilde \by_{t}$ plays a critical role in linking individual model forecasts to the observable variable. However, this relationship remains susceptible to misspecification risks, particularly when the candidate models may be false. To address this limitation, \cite{B:13} employed a parametric latent variable, which serves as a hidden process of the time-varying weights. Before going further to detail the latent process related to the weights in \cite{B:13}, we first state the general framework of the combination forecast. Denote $W_t=(\bw_{t}^{1},\bw_{t}^{2},\cdots,\bw_{t}^{L})\in\R^{K\times L}$ be the time-varying weights of the combination scheme at time $t$, with $\bw_{t}^{l} = (w_{1,t}^{l},w_{2,t}^l,\cdots,w_{K,t}^{l})'\in\Delta_{[0,1]^K}$, being the weights assigned to the $l$-th component of each predictor, $l=1,\cdots,L$. In this context, $\Delta_{[0,1]^K}\subset\R^K$ represents the collection of vectors with non-negative components that sum to $1$, known as the standard $K$-dimensional simplex. It is evident that $W_t \in \W$ represents a $K \times L$ matrix, where each column sums to $1$.

Referring again to Equation~\eqref{eqn-2.1}, the first multiplier in the integral $p(\by_t|\tilde \by_{1:t},\by_{1:t-1})$ represents the impact of all predictors up to time $t$ as well as the historical data of the observables on the observable variable at time $t$. With the weight matrix $W_t$ on hand, it is easy to derive that 
\begin{align}\label{eqn-2.3}\notag
    p(\by_t|\tilde \by_{1:t},\by_{1:t-1})
    =&\int_\W p(\by_t,W_t|\tilde \by_{1:t},\by_{1:t-1})dW_t\\
    =&\int_\W p(\by_t|W_t,\tilde \by_{1:t},\by_{1:t-1})p(W_t|\tilde \by_{1:t},\by_{1:t-1})dW_t.
\end{align}
It is assumed in \cite{B:13} that 
\begin{enumerate}
    \item[(1)] the density estimation of $\by_t$ depends only on $W_t$ and $\tilde \by_t$, i.e., \linebreak
    \begin{equation}\label{eqn-2.4}
        p(\by_t|W_t,\tilde \by_{1:t},\by_{1:t-1})=p(\by_t|W_t,\tilde \by_t).
    \end{equation}
    \item[(2)] the prior of $W_t$ is independent of $\tilde \by_t$, i.e., 
    \begin{equation}\label{eqn-2.5}
        p(W_t|\tilde \by_{1:t},\by_{1:t-1})=p(W_t|\tilde \by_{1:t-1},\by_{1:t-1}).
    \end{equation}
\end{enumerate}
Under these assumptions, Equation~\eqref{eqn-2.3} is simplified as
\begin{equation}\label{eqn-2.6}
    p(\by_t|\tilde \by_{1:t},\by_{1:t-1})
    =\int_\W p(\by_t|W_t,\tilde \by_t)p(W_t|\tilde \by_{1:t-1},\by_{1:t-1})dW_t.
\end{equation}

In \cite{B:13}, the transition density of the weights is assumed to be a first-order Markovian dynamics  $p(W_t|W_{t-1},\tilde \by_{1:t-1},\by_{1:t-1})$ and depends on the observables' history $\by_{1:t-1}$ and those of the predictors $\tilde\by_{1:t-1}$. Thus, the second term in the integrand Equation~\eqref{eqn-2.6} is written as
\begin{align}\label{eqn-2.7}\notag
     &p(W_t|\tilde\by_{1:t-1},\by_{1:t-1})\\
    =&\int_\W p(W_t|W_{t-1},\tilde\by_{1:t-1},\by_{1:t-1})p(W_{t-1}|\tilde\by_{1:t-1},\by_{1:t-1})dW_{t-1}.
\end{align}
Here, $p(W_{t-1}|\tilde\by_{1:t-1},\by_{1:t-1})$ is the posterior density of the weights at time $t-1$. Within the Bayesian framework, the posterior density of the weights \linebreak $p(W_{t-1}|\by_{1:t-1},\tilde\by_{1:t-1})$ in Equation~\eqref{eqn-2.7} is updated by the Bayesian formula:
\begin{align}\label{eqn-2.8}\notag
    p(W_{t-1}|\tilde\by_{1:t-1},\by_{1:t-1})
    =& \frac{p(\by_{t-1}|W_{t-1},\tilde\by_{1:t-1},\by_{1:t-2})p(W_{t-1}|\tilde\by_{1:t-1},\by_{1:t-2})}{p(\by_{t-1}|\tilde\by_{1:t-1},\by_{1:t-2})}\\\notag
    =&\frac{p(\by_{t-1}|W_{t-1},\tilde\by_{t-1})p(W_{t-1}|\tilde\by_{1:t-2},\by_{1:t-2})}{p(\by_{t-1}|\tilde\by_{1:t-1},\by_{1:t-2})}\\
    \propto&p(\by_{t-1}|W_{t-1},\tilde\by_{t-1})p(W_{t-1}|\tilde\by_{1:t-2},\by_{1:t-2}),
\end{align}
by Equation~\eqref{eqn-2.4}-\eqref{eqn-2.5}. Here, $p(W_{t-1}|\tilde \by_{1:t-2},\by_{1:t-2})$ is the prior density of the weights at time $t-1$. The prior and posterior densities of the weights exhibit a recursive progression over time.   

\begin{figure}[!ht]
 \centering
    \includegraphics[trim=45 320 45 80, clip,width=\textwidth]{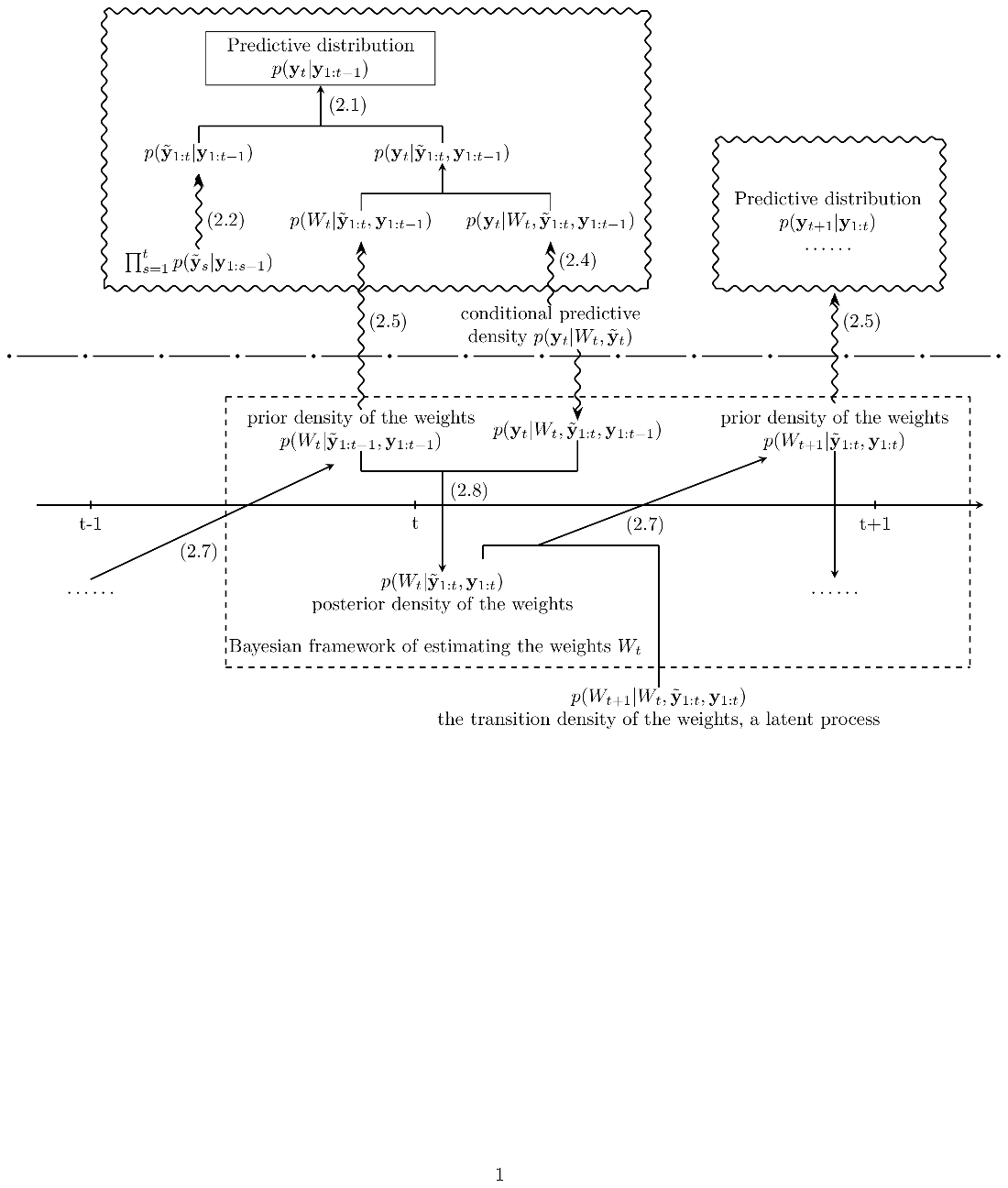}
\caption{The flowchart of the TVW proposed in \cite{B:13}}
\label{fig-1}
\end{figure}

For clarity, we summarize this framework in Figure \ref{fig-1}. We emphasize that the Bayesian idea is reflected in the way we update the weights $W_t$ based on new coming observations $\by_t$.

To be practical, one needs to specify the likelihood density $p(\by_t|W_t,\tilde\by_t)$ in Equation~\eqref{eqn-2.6} and the latent weights' transition density $p(W_t|W_{t-1},\tilde\by_{1:t-1},\by_{1:t-1})$ in Equation~\eqref{eqn-2.7}. This reflects this framework's flexibility. In \cite{B:13}, the likelihood density is assumed to be Gaussian density, i.e.
\begin{equation}\label{eqn-2.9}
    p(\by_t|W_t,\tilde\by_t)\propto|\Sigma|^{-\frac12}\exp\left\{-\frac12\left[\by_t-\diag(W_t\tilde\by_t)\right]'\Sigma^{-1}\left[\by_t-\diag(W_t\tilde\by_t)\right]\right\},
\end{equation}
where $\Sigma$ is the covariance matrix, $|A|$ and $\diag(A)$ are the determinant and the diagonal entries of the matrix $A$, respectively. For the weights' transition density, \cite{B:13} proposed a latent variable $X_t=\left(\bx_t^1,\bx_t^2, \cdots, \bx_t^L\right)\in\R^{K\times L}$, where $ \bx_t^l=\left(x_{1, t}^l, x_{2, t}^l, \cdots, x_{K, t}^l\right)'\in \R^K$, $l=1,\cdots,L$, governing the regime-specific weights. Let $\bx_t:=\bvec(X_t)\in\R^{KL\times 1}$, where $\bvec(A)$ is the vectorization of the matrix $A$ column by column. In \cite{B:13}, it is assumed that 
\begin{equation}\label{2.10} 
    p\left(\bx_t\mid\bx_{t-1}\right) 
    \propto |\Lambda|^{-\frac{1}{2}} \exp \left\{-\frac{1}{2}\left(\bx_t-\bx_{t-1}\right)^{\prime} \Lambda^{-1}\left(\bx_t-\bx_{t-1}\right)\right\},
\end{equation}
where $\Lambda$ is the covariance matrix. The weights $\bw^l_t=(w_{1,t}^l,\cdots,w_{K,t}^l)$ is obtained by the logistic transformation, i.e.
\begin{equation}\label{2.11}
    w_{k,t}^l:=g_k(x_{k,t}^l)
    =\frac{\exp\left(x_{k,t}^{l}\right)}{\sum_{k=1}^{K}\exp\left(x_{k,t}^{l}\right)},
\end{equation} 
for $l=1,\cdots,L$, where $\bg(x):=(g_1(x),\cdots,g_K(x))$ is a flexible link function of $x$. Here, we assume it to be a softmax function. This transformation guarantees that $\bw_t^l\in\Delta_{[0,1]^{K}}$ is indeed weights assigned to each model. 



Additionally, \cite{B:13} introduced a learning mechanism in the weight structure, using the exponentially weighted moving average of historical forecast errors as a learning strategy to refine the model. Unfortunately, the performance of this learning mechanism is insignificant in the empirical experiments. We believe that the diversity of the predictors, introduced in Section \ref{sec-2.2} below, is a better feedback mechanism to adjust the weight, enriching the dynamic structure of combination forecast.

\subsection{Model diversity}\label{sec-2.2}

\cite{LW:20} emphasized that diversity among the individual methods being combined is an essential factor for superior performance in exploring successful combinations in the M4 competition. The reduction in relative error in combination forecast is influenced by the degree to which the constituent models incorporate unique or independent information, see \cite{T:19}. While much of the existing literature has mentioned the importance of diversity or qualitatively analyzes it post-combination, \cite{KW:22} defined the pairwise error correlations among those methods as the diversity. To ensure comparability of diversity across time series with different scales, they normalized the diversity by averaging over all pairs of individual methods. Here, we define a similar concept, the scaled diversity, i.e.,
\begin{align}\label{div}
    \divg^l_{k,t,h}:=\frac{\sum\limits_{i=1}^{K}(\tilde\by^l_{k,t+h}-\tilde\by^l_{i,t+h})^2}{\sum\limits_{i,j=1}^{K}(\tilde\by^l_{i,t+h}-\tilde\by^l_{j,t+h})^2}, 
\end{align}
for $l=1,\cdots,L$, $k=1,2,\cdots,K$, $h$ is the forecast horizon. It should be noted that the diversity at time $t$ $\Divg_{t,h}$ is a $K\times L$ matrix with its $(k,l)$-th component being $\divg^l_{k,t,h}$. Significantly, it is independent of both the history of observables $\by_{1:t - 1}$ and the past estimations of individual methods $\tilde\by_{1:t - 1}$. This characteristic endows the diversity with the capacity to offer a robust weight adjustment mechanism when historical data are inadequate or scarce. 

In Section \ref{sec-3}, we explore a dynamic weighting process that synchronizes the latent process $\bx_t$ in Equation~\eqref{2.10} with the scaled diversity defined in Equation~\eqref{div}. The aim is to enhance the performance of combination forecast without relying on any additional information, except for the historical observables and the $K$ predictors. The linchpin of this approach is to systematically incorporate the information encapsulated within the diversity into the weight-determination mechanism.

\subsection{Other combination models}\label{sec-2.3}
For comparisons in Section \ref{sec-5}, besides the TVW introduced in Section \ref{sec-2.1}, we briefly mention the Equal weighting (Equal) and BMA here. 

Equal assigns equal weights $w_{k,t}=1/K$ to all candidate models, $k=1,\cdots,K$. Such parsimonious weighting is commonly used and is sometimes found to outperform more sophisticated adaptive combination schemes. This is called ``forecast combination puzzle", see \cite{CP:18, CMVW:16,T:06}.

BMA addresses model uncertainty by combining predictions from multiple candidate models, weighted by their posterior probabilities. This approach has been applied in the econometrics literature by \cite{AG:07,HM:07}. \cite{JMV:10} computed the recursive weights by
$$
w_{k,t}=\frac{\exp\left \{\sum\limits_{t=\underline{t}}^{\overline{t}} \ln p(\tilde\by_{k,t}|\tilde\by_{1:t-1},\by_{1:t-1})\right\}}
{\sum\limits_{k=1}^{K} \exp\left\{\sum\limits_{t=\underline{t}}^{\overline{t}} \ln p(\tilde\by_{k,t}|\tilde\by_{1:t-1},\by_{1:t-1})\right\}},
$$
$k=1,\cdots,K$, where $\bar{t}$ and $\underline{t}$ denote the beginning and the end of the evaluation period, respectively. In Section \ref{sec-5}, to contrast ex-post stability with real-time adaptability against structural breaks, we implement two BMA variants: the full-sample weights and 24-month rolling-window updated weights, denoted as BMA and BMA\_roll, respectively.

\section{Forward-looking feedback framework and forecast combination with diversity-driven time-varying weights}\label{sec-3}

\setcounter{equation}{0}

In this section, we shall propose a forward-looking feedback framework under the assumption that 
\begin{enumerate}
  \item[(0)]\quad  At time $t$, the $K$ predictors are capable of making $h$-step ahead predictions, denoted as $\tilde\by_{t:t+h}$, for some $h=1,2,\cdots$. 
\end{enumerate}
We believe that these predictions provide extra information to yield better estimate for the current weights $W_t$, even if the models are misspecified. In Section \ref{sec-3.1}, we re-establish the framework to infuse $\tilde\by_{t:t+h}$ into the weight estimation. In Section \ref{sec-3.2}, we propose an algorithm to specifically make the forward-looking framework implementable. The diversity $\Divg_{t,h}$ is introduced to reflect the information from $\tilde\by_{t:t+h}$. We call it forecast combination with diversity-driven time-varying weight (DTVW) in the sequel.

\subsection{The forward-looking feedback framework}\label{sec-3.1}

As in Section \ref{sec-2.1}, we aim to obtain the predictive distribution $p(\by_t|\by_{1:t-1})$. By Assumption (0), we have $\tilde\by_{1:t+h-1}$ available at time $t-1$, but only with the past observable variables $\by_{1:t-1}$ at hand. Thus, we have
\begin{equation}\label{eqn-3.1}
    p(\by_t|\by_{1:t-1})=\int_\Y p(\by_t|\tilde\by_{1:t+h-1},\by_{1:t-1})p(\tilde\by_{1:t+h-1},\by_{1:t-1})d\tilde\by_{1:t+h-1},
\end{equation}
where 
\begin{equation}\label{eqn-3.2}
    p(\tilde\by_{1:t+h-1}|\by_{1:t-1})
    =\prod\limits_{s=1}^{t}p(\tilde\by_{s+h-1}|\by_{1:s-1}).
\end{equation}
Similarly to Equation~\eqref{eqn-2.3}, the weight matrix $W_t$ plays a vital role in the predictive distribution.
\begin{align}\label{eqn-3.3}\notag
    p(\by_t|\tilde \by_{1:t+h-1},\by_{1:t-1})
    =&\int_\W p(\by_t|W_t,\tilde \by_{1:t+h-1},\by_{1:t-1})p(W_t|\tilde \by_{1:t+h-1},\by_{1:t-1})dW_t\\
    =&\int_\W p(\by_t|W_t,\tilde \by_{t:t+h-1})p(W_t|\tilde \by_{1:t+h-1},\by_{1:t-1})dW_t,
\end{align}
if the following two assumptions, corresponding to (1)-(2) in Section \ref{sec-2.1}, hold: 
\begin{enumerate}
    \item[(1)]\quad the density of $\by_t$ depends only on $W_t$ and $\tilde\by_{t:t+h}$, i.e.,
    \begin{equation}\label{eqn-3.4}
        p(\by_t|W_t,\tilde\by_{1:t+h-1},\by_{1:t-1})=p(\by_t|W_t,\tilde\by_{t:t+h-1}).
    \end{equation}
    \item[(2)$'$]\quad the prior density of the weight $W_t$ depends on $\tilde\by_{1:t+h-1}$, but not those predictions beyond $t+h$, i.e., 
    \begin{equation}\label{eqn-3.5}
        p(W_t|\tilde\by_{1:t+h},\by_{1:t-1})=p(W_t|\tilde\by_{1:t+h-1},\by_{1:t-1}).
    \end{equation}
\end{enumerate}
We make some remarks about these two assumptions. Assumption (1) is exactly the same as the previous one. About Assumption (2)$'$, the primary difference from before is that we make use of the forward-looking information provided by $\tilde\by_{t:t+h-1}$ to estimate $W_t$. Note that the important fact here is that when we make a prior estimate of the weight at time $t-1$, the $K$ predictors $\tilde\by_{t:t+h-1}$ at time $t$ have already been available by Equation~\eqref{eqn-3.2}. 

The first multiplier in the integral on the right-hand side of Equation~\eqref{eqn-3.3} is the likelihood density, while the second one is the prior distribution of the weight $W_t$. Similarly as before, the prior and posterior densities revolve in a cycle:
\begin{align}\label{eqn-3.6}\notag
    &p(W_t|\tilde\by_{1:t+h-1},\by_{1:t-1})\\
    =&\int_\W p(W_t|W_{t-1},\tilde\by_{1:t+h-1},\by_{1:t-1})p(W_{t-1}|\tilde\by_{1:t+h-1},\by_{1:t-1})dW_{t-1},
\end{align}
and 
\begin{align}\label{eqn-3.7}\notag
  p(W_{t-1}|\tilde\by_{1:t+h-1},\by_{1:t-1})
    \propto&p(\by_{t-1}|W_{t-1},\tilde\by_{1:t+h-1},\by_{1:t-2})p(W_{t-1}|\tilde\by_{1:t+h-1},\by_{1:t-2})\\
    \propto& p(\by_{t-1}|W_{t-1},\tilde\by_{t-1:t+h-1})p(W_{t-1}|\tilde\by_{1:t+h-2},\by_{1:t-2}),    
\end{align}
by Assumption (1)-(2)$'$.

\begin{figure}[!ht]
 \centering
    \includegraphics[trim=15 200 0 0, clip,width=\textwidth]{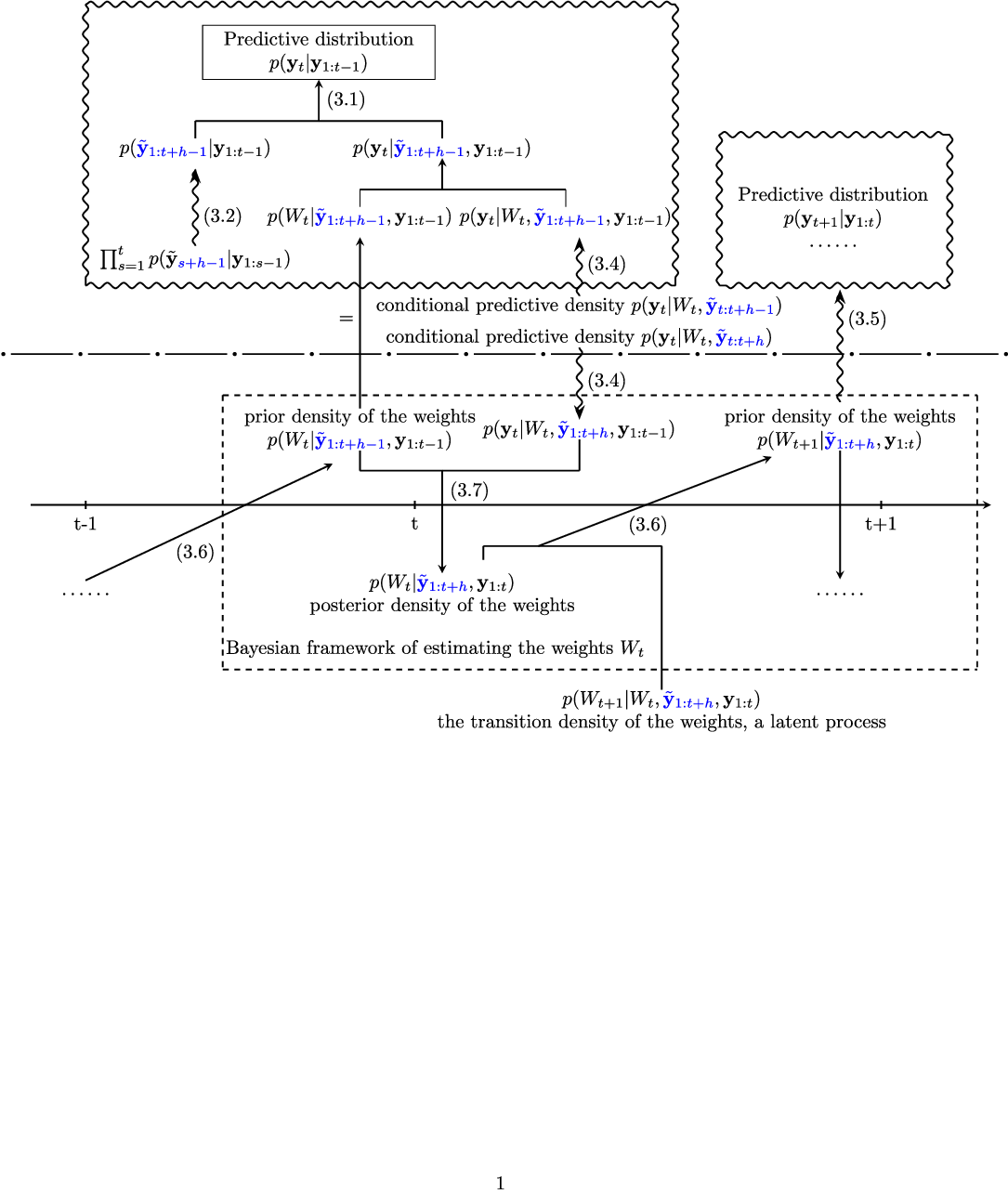}
\caption{The flowchart of the forward-looking framework}
\label{fig-2}
\end{figure}

The flowchart of the forward-looking feedback framework is presented in Figure \ref{fig-2}. The blue fonts therein emphasize the differences from those in Figure \ref{fig-1}. To be practical, we specify the likelihood density $p(\by_t|W_t,\tilde\by_{t:t+h-1})$ in Equation~\eqref{eqn-3.3} similar as Equation~\eqref{eqn-2.9}.
This new framework offers several advantages:
\begin{enumerate}
   \item[\textbullet] 
   Predictions from different models for time instants beyond $t$ may possess partial information of the observable variables $\mathbf{y}_t$. These predictions potentially offer more accurate forecasts of the future behavior of $\mathbf{y}_t$. This framework elucidates the appropriate parts at which these predictions can be effectively incorporated.
     \item[\textbullet] The transition density of the weights $p(W_{t+1}|W_{t},\tilde\by_{1:t+h},\by_{1:t})$ furnishes the prior distribution for the time instant $t + 1$. The multi-step ahead predictions $\tilde\by_{t:t+h}$ can potentially yield a more refined prior distribution compared to the pure Brownian motion described in Equation~\eqref{2.10}. The latter, in fact, contains no relevant information whatsoever. An apt construction of this latent process could achieve a significant breakthrough.
\end{enumerate}

\subsection{Forecast combination with diversity-driven time-varying weight (DTVW)}\label{sec-3.2}

In this section, we propose a latent process for the transition density in Section \ref{sec-3.2.1}, which serves to synchronize the scaled diversity in Equation~\eqref{div}. As expounded in Section \ref{sec-2.2}, the scaled diversity is a time-varying variable. It furnishes real-time feedback regarding the disparities among multi-step ahead forecasts derived from individual models. This diversity has been scaled to a dimensionless quantity, and our objective is to integrate it into the driver process of the weights. Subsequently, this method is designated as forecast combination with DTVW. For its practical implementation, the Bayesian framework enclosed within the dashed box in Figure \ref {fig-2} precisely pertains to the domain of nonlinear filtering. We elaborate in detail in Section \ref{sec-3.2.2} on how the particle filter (PF), also known as the sequential Monte Carlo method in statistics, is employed to approximate the prior and posterior densities, as well as the predictive distribution.

\subsubsection{The latent process incorporated with diversity}\label{sec-3.2.1}

To integrate the diversity into the latent process, we use the regression model
\begin{equation}\label{eqn-3.8}
    \bx_t=\theta_{0,t}+\theta_{1,t}\bx_{t-1}+\theta_{2,t}\mathbf{div}_{t,h}+\bepsilon_{1,t},
\end{equation}
where  $\bx_t=\bvec{(X_t)}\in\R^{KL\times 1}$ and  ${\mathbf{div}_{t,h}}=\bvec(\Divg_{t,h}')\in\R^{KL\times1}$, and $\bepsilon_{1,t}\sim\mathcal{N}(0,\Sigma_{1,t})$. $\theta_{0,t}$ is the intercept and the regression coefficients $(\theta_{1,t},\theta_{2,t})$ balance the effects from $\bx_{t-1}$ and the scaled diversity $\Divg_{t,h}$. Similar to $\bx_t$ in Section \ref{sec-2.1}, $\btheta_t=(\theta_{0,t},\theta_{1,t},\theta_{2,t})\in[-1,1]^3$ is also within the Bayesian framework, which is subject to prediction and update at each time instant.

The coefficient $\btheta_{t}$ is scaled into $[-1,1]^3$ by linking another latent process $\balpha_{t}=(\alpha_{0,t},\alpha_{1,t},\alpha_{2,t})\in\R^3$ with 
\begin{equation}\label{eqn-3.9}
     \theta_{i,t}= 2\left( \frac{1}{1 + e^{-\alpha_{i,t}}} - \frac{1}{2} \right),
\end{equation}
where $\balpha_{t}$ is driven by a Brownian motion \begin{equation}\label{eqn-3.10}
    \balpha_{t}=\balpha_{t-1}+\bepsilon_{2,t},
\end{equation}
where $\bepsilon_{2,t}\sim\mathcal{N}({\bm0,\Sigma_{2,t}})$. 

The transformation for $\theta_{i,t}$ in Equation~\eqref{eqn-3.9} serves to scale it to a range comparable to that of the diversity measure, which lies within $[0,1]$. Other scaling functions could also be employed for this purpose.

As in Equation~\eqref{2.11}, the logistic transformation of $x_{k,t}^l$ is the weight of the $l$-th variable in the $k$-th model.

The proposed latent process described by Equation~\eqref{eqn-3.8} and~\eqref{eqn-3.10} reveals that the latent variable 
$\bx_t$ at time $t$ is influenced not only by its previous value $\bx_{t-1}$, but also by the h-step ahead predictions of individual models. Through this approach, information from the future acts as a feedback mechanism, seamlessly integrated into the dynamic framework governing the weights.

\subsubsection{The DTVW}\label{sec-3.2.2}

With the forward-looking feedback framework proposed in Section \ref{sec-3.1}, we detail the DTVW combination forecast below.

The predictive distribution $p(\by_t|\by_{1:t-1})$ is given by Equation~\eqref{eqn-3.1}-\eqref{eqn-3.3}, if one obtains the prior density of the weights $p(W_t|\tilde\by_{1:t+h-1},\by_{1:t-1})$, equivalently $p(\bx_t|\tilde\by_{1:t+h-1},\by_{1:t-1})$, since from our construction Equation~\eqref{2.11}, the weight $\bw_t=\bvec{(W_t)}$ is completely determined by $\bx_t$. Here, we detail how to get it with the latent process proposed in Equation~\eqref{eqn-3.8} and~\eqref{eqn-3.10}. Let us denote $\bv_t:=(\bx_t,\balpha_t)\in\R^{KL+3}$. The prior distribution of $\bx_t$ can be easily obtained by integrating $\balpha_t$ out from the prior distribution of $\bv_t$:
\begin{equation}\label{eqn-3.11}
\begin{aligned}
     p(\bx_t|\tilde\by_{1:t+h-1},\by_{1:t-1})
     &=\int_{\R^3} p(\bx_t,\balpha_t|\tilde\by_{1:t+h-1},\by_{1:t-1})d\balpha_t \\
     &=\int_{\R^3} p(\bv_t|\tilde\by_{1:t+h-1},\by_{1:t-1})d\balpha_t.
\end{aligned}
\end{equation}

Thus, it is sufficient to obtain the prior distribution of $\bv_t$. From the dashed box in Figure \ref{fig-2}, the weight $W_t$ is within the Bayesian framework to be predicted and updated at each instant, so does $\bv_t$. Two assumptions, similar as Assumption (1)-(2)', are proposed:
\begin{enumerate}
    \item[(1)$'$]\quad the density of $\by_t$ depends only on $W_t$ and $\tilde\by_t$, i.e.
    \begin{equation}\label{eqn-3.26}
        p(\by_t|\bv_t,\tilde\by_{1:t+h-1+m},\by_{1:t-1})=p(\by_t|W_t,\tilde\by_t),
    \end{equation}
    for any $m=0,1,2,\cdots$.
    \item[(2)$''$]\quad the prior distribution of $\bv_t$ depends on $\tilde\by_{1:t+h-1}$ at time $t-1$, but not those predictions beyond $t+h$, i.e. 
    \begin{equation}\label{eqn-3.13}
        p(\bv_t|\tilde\by_{1:t+h+m},\by_{1:t-1})=p(\bv_t|\tilde\by_{1:t+h-1},\by_{1:t-1}),
    \end{equation}
for any $m=0,1,\cdots$.
\end{enumerate}

The prior distribution of $\bv_t$ is
     \begin{align}\label{eqn-3.17}
     &p(\bv_t|\tilde\by_{1:t+h-1},\by_{1:t-1}) \\\notag
      =&\int_{\R^{KL+3}}p(\bv_t|\bv_{t-1},\tilde\by_{1:t+h-1},\by_{1:t-1})p(\bv_{t-1}|\tilde\by_{1:t+h-1},\by_{1:t-1})d\bv_{t-1},
\end{align}
where the second multiplier in the integrand is the posterior distribution of $\bv_{t-1}$ to be discussed later in Equation~\eqref{eqn-3.16}, and the first one is described by the latent process Equation~\eqref{eqn-3.8} and~\eqref{eqn-3.10}:
\begin{align}\label{eqn-3.12}\notag
    &p(\bv_{t}|\bv_{t-1},\tilde\by_{1:t+h-1},\by_{1:t-1})\\\notag
    =&p(\bx_{t}|\balpha_{t},\bv_{t-1},\tilde\by_{1:t+h-1},\by_{1:t-1})p(\balpha_{t}|\bv_{t-1},\tilde\by_{1:t+h-1},\by_{1:t-1})\\\notag
    =&p(\bx_{t}|\balpha_{t},\bx_{t-1},{\mathbf{div}_{t,h}})p(\balpha_{t}|\balpha_{t-1})\\
    \propto&|\Sigma_{1,t}|^{-\frac{1}{2}} \exp \left\{-\frac{1}{2} \left( \bx_{t}-\theta_{0,t}-\theta_{1,t}\bx_{t-1}  -\theta_{2,t}\mathbf{div}_{t,h}\right)'\Sigma_{1,t}^{-1
    }\right.\\\notag
        &\phantom{|\Sigma_{1,t}|^{-\frac{1}{2}}aaaaaa}\cdot\left.\left(\bx_{t}-\theta_{0,t}-\theta_{1,t}\bx_{t-1} -\theta_{2,t}\mathbf{div}_{t,h} \right) \right\}\\\notag
        &\cdot|\Sigma_{2,t}|^{-\frac{1}{2}} \exp \left\{-\frac{1}{2}\left(\balpha_{t}-\balpha_{t-1}\right)'\Sigma_{2,t}^{-1}\left(\balpha_{t}-\balpha_{t-1}\right)\right\},
\end{align}
where $\btheta_t$ is linked to $\balpha_t$ deterministically by Equation~\eqref{eqn-3.9}. The second equality in Equation~\eqref{eqn-3.12} reflects the role of the latent processes Equation~\eqref{eqn-3.8} and~\eqref{eqn-3.10} in the DTVW. The posterior distribution $p(\bv_{t-1}|\tilde \by_{1:t+h-1},\by_{1:t-1})$ is updated according to
\begin{align}\label{eqn-3.16}\notag
   p(\bv_{t-1}|\tilde\by_{1:t+h-1},\by_{1:t-1})
   \propto& p(\by_{t-1}|\bv_{t-1},\tilde\by_{1:t+h-1},\by_{1:t-2})p(\bv_{t-1}|\tilde\by_{1:t+h-1},\by_{1:t-2})\\\notag
    =& p(\by_{t-1}|\bx_{t-1},\tilde\by_{t-1})p(\bv_{t-1}|\tilde\by_{1:t+h-2},\by_{1:t-2})\\
    \overset{\eqref{eqn-2.9}}\propto&|\Sigma|^{-\frac12}\exp\left\{-\frac12\left[\by_{t-1}-\diag(W_{t-1}\tilde\by_{t-1})\right]'\Sigma^{-1}\right.\\\notag
    &\phantom{|\Sigma|^{-\frac12}\exp aa}\left.\cdot\left[\by_{t-1}-\diag(W_{t-1}\tilde\by_{t-1})\right]\right\}\\\notag
        &\cdot p(\bv_{t-1}|\tilde\by_{1:t+h-2},\by_{1:t-2}),
\end{align}
by Assumption (1)$'$-(2)$''$, where the weight $W_t$ is the logistic transformation of $\bx_t$ Equation~\eqref{2.11} and the prior distribution of $\bv_{t-1}$ is as that for $\bv_t$ in Equation~\eqref{eqn-3.17}. 

The principal distinctions between TVW and DTVW are twofold. First, in terms of the construction of the latent process $\bx_t$, the DTVW's latent process not only synchronizes the historical information encapsulated in $\bx_{t - 1}$ but also incorporates the predictions from individual predictors, $\Divg_{t,h}$. Here, the diversity is regarded as a crucial feature, significantly enhancing the performance of the combination forecast. Second, the regression coefficients $\btheta_t\in\mathbb{R}^3$, as introduced in Equation~\eqref{eqn-3.8}, play a pivotal role. They offer additional degrees of freedom, enabling a more refined balance between the influence of historical data and future-oriented information.

\subsubsection{The implementation of DTVW using particle filter}\label{sec-3.2.3}

From Section \ref{sec-3.2.1}, the DTVW inherently involves estimating the prior and posterior distributions of $\bv_t$, Equation~\eqref{eqn-3.17} and~\eqref{eqn-3.16}, and the prior distribution of the weights $W_t$ is subsequently obtained by Equation~\eqref{eqn-3.11}. This task lies exactly within the domain of nonlinear filtering (NLF). In this subsection, we systematically elaborate on the application of particle filtering (PF) for approximating these distributions. Notably, PF represents just one of the approaches within the domain of NLF for carrying out this estimation. Other alternatives include, but are not restricted to, the extended Kalman filter, unscented Kalman filter, ensemble Kalman filter, among others. A comparative study on the performance of different NLF techniques applied to combination forecast is one of our on-going researches.

As the analytical solution of the prior and posterior distributions cannot be generally known for the nonlinear state space models, the numerical approximation methods are inevitable. The PF is essentially using the empirical distribution of the i.i.d. samples to represent the distributions, which asserts to be optimal~\citep{DFG:01}. To be more specific, let $\chi_t:=\{\bv_t^i,\omega_t^i\}_{i=1}^{N}$, where $\bv_t^i$ is the $i$th sample at time $t$ paired with its weight $\omega_t^i$, and the sample size is $N$. The particles $\bv_t^i$ as well as its weight $\omega_t^i$ are predicted and updated according to Equation~\eqref{eqn-3.8},~\eqref{eqn-3.10} and~\eqref{eqn-3.16}, respectively. The procedure is summarized in Algorithm \ref{algorithm-pf}. It is easy to see that the empirical prior and posterior distributions in Equation~\eqref{eqn-3.17} and~\eqref{eqn-3.16} are
\begin{align*}
    p_{N}(\bv_t|\tilde \by_{1:t+h-1},\by_{1:t-1})
    =&\sum_{i=1}^{N}p(\bv_t|\bv_{t-1},\tilde \by_{1:t+h-1},\by_{1:t-1})\omega_{t-1}^i\delta_{\bv_{t-1}^i}(\bv_{t-1}),\\
    p_{N}(\bv_t|\tilde\by_{1:t+h-1},\by_{1:t})
    =&\sum_{i=1}^{N}\omega_{t}^i\delta_{\bv_{t}^i}(\bv_t),
\end{align*}
respectively, where $\omega_t^i\propto\omega_{t-1}^i p(\by_t|W_t^i,\tilde\by_t)$, and $\delta_x(y)$ is the Dirac mass centered at $x$.

\begin{algorithm}[h!]
\caption{Weight estimation and prediction of $\by_t$ using particle filter}
\begin{algorithmic}[1]\label{algorithm-pf}
  \STATE \% Initialization
  \FOR{$i = 1$ to $N$}

      \STATE Sample $\bv_{0}^i\overset{i.i.d.}\sim p(\bv_{0})$ pairing with the weight $\omega_0^{i} = \frac{1}{N}$, where $\bv_0^i:=(\bx_0^i,\balpha_0^i)\in\R^{KL+3}$.
\ENDFOR

  \STATE \% The particle filter
  \FOR{$t = 1$ to $T$}
    \FOR{$i = 1$ to $N$}
          \STATE \% Predict  
          \STATE The $i$th particle $\bv_{t-1}^{i}$ is propagated by Equation~\eqref{eqn-3.8} and~\eqref{eqn-3.10} to obtain $\bv_{t}^i$, where $\btheta_t^i$ is linked with $\balpha_t^i$ by Equation~\eqref{eqn-3.9}.
             
      \STATE \% Update
      \STATE The weight $\omega_t^{i} \propto \omega_{t-1}^{i}p(\by_{t}|W_t^{i},\tilde\by_{t})$, where $W_t^i$ is the component-wise logistic transformation of $X_t^i$ Equation~\eqref{2.11}.
    \ENDFOR
\STATE \% Resampling
    \STATE Normalize weights: $\omega_t^{i} \leftarrow \frac{\omega_t^{i}}{\sum_j \omega_t^{j}}$, and then compute the effective sample size (ESS): $\mathrm{ESS}_t = \frac{1}{N\sum_i (\omega_t^{i})^2}$.
    \IF{$\mathrm{ESS}_t < \kappa$, a threshold,}
      \STATE Resample particles $\{\bv_t^{i}\}_{i=1}^{N}$ according to the weights $\omega_t^{i}$, and then reset them to be equal, i.e. $\omega_t^{(i)} = \frac{1}{N}$.
    \ENDIF
    \STATE Predict $\by_t$ by $\bar{W}_t\tilde\by_t$, where $\bar{W}_t=\frac1N\sum_{i=1}^NW_t^i$.
  \ENDFOR
\end{algorithmic}
\end{algorithm}

\section{Monte Carlo evidences}\label{sec-4}
\setcounter{equation}{0}

\subsection{Evaluation Metrics}\label{sec-4.1}

To assess the predictive performance of the individual models/combination forecast methods, we shall evaluate them by commonly-used evaluation metrics for both point and density forecasts, such as Root Mean Squared Forecast Errors (RMSFE) for point forecasts,  Logarithmic Score (LS) and Continuous Rank Probability Score (CRPS) for density forecasts.

The RMSFE is defined as
\[
    \mathrm{RMSFE}=\sqrt{\frac{1}{\bar{t}-\underline{t}+1}\sum_{t=\underline{t}}^{\bar{t}}\left(\mathbf{y}_t-\tilde{\mathbf{y}}_t\right)^2},
\]
where $\bar{t}$ and $\underline{t}$ are the beginning and the end of the evaluation period, $\tilde\by_t$ is the prediction of the observation $\by_t$ at time $t$. This metric measures the average dispersion between predictions and observations. The smaller the RMSFE is, the more accurate the forecast is.

The LS is defined as
\[
    \mathrm{LS}=-\frac{1}{\bar{t}-\underline{t}+1}\sum_{t=\underline{t}}^{\overline{t}}\ln p({\mathbf{y}}_{t+1}|\mathbf{y}_{1:t}),
\]
where $p({\mathbf{y}}_{t+1}|\mathbf{y}_{1:t})$ denotes the predictive distribution of $\by_{t+1}$ conditioned on the historical observations $\by_{1:t}$. The LS evaluates calibration quality by assessing the average negative log-likelihood assigned to actual outcomes, applying a strict proper scoring rule that penalizes poor distributional calibration. Lower LS values imply better probabilistic forecasts, since they indicate higher predictive densities at the realized values.

The CRPS is defined as 
\begin{align*}
    \mathrm{CRPS}(F_{t+1},\by_{t+1})
    = &\int\left|F_{t+1}(z)-1_{\{z\geq\mathbf{y}_{t+1}\}}(z)\right|^2dz\\
    = &\mathbb{E}_{t+1}|\tilde\by_{t+1}-\by_{t+1}| - \frac{1}{2}\mathbb{E}_{t+1}|\tilde\by_{t+1}-\tilde\by_{t+1}'|,
\end{align*}
where $F_{t+1}$ is the cumulative distribution function (CDF) of the predictive distribution $p(\by_{t+1}|\by_{1:t})$, $\E_{t+1}$ is the expectation with respect to $F_{t+1}$, $\tilde\by_{t+1}$ and $\tilde\by_{t+1}'$ are independent copies of predictions from the predictive distribution, and $\by_{t+1}$ is the actual observation. The CRPS is a metric, which measures the difference between the predictive CDF and the step function representing the actual outcome. It provides a combined assessment of the accuracy of both the point prediction and the uncertainty represented by the forecast distribution, see Section 4.2~\citep{GR:07}. A lower CRPS indicates better predictive performance.

\subsection{Model validation}\label{sec-4.2}

In this subsection, we shall examine whether our DTVW improves the TVW in the predictive performance, as well as the true model selection capability. To investigate this, we experiment two types of model sets: the simple complete and the complex incomplete ones. 

\subsubsection{Simple complete model sets}\label{sec-4.2.1}

It is shown in \cite{B:13} that the true model selection capability of the TVW is superior in the complete model sets. To see whether we can recover the TVW from the DTVW by letting $\theta_{2,t}\equiv0$, for all $t$. That is, we shall only estimate $\theta_{0,t}$ and $\theta_{1,t}$ in Equation~\eqref{eqn-3.8} without incorporating diversity, to see whether the first two parameters are adaptively estimated to be the same as those in the TVW~\citep{B:13}, i.e. $\theta_{0,t}\equiv0$, $\theta_{1,t}\equiv1$. Let us call this scheme adaptive TVW. Furthermore, we shall include the diversity-related parameter $\theta_{2,t}$ in the estimation to assess whether accounting for diversity can further improve predictive performance. During the implementation of these experiments, we estimate $\alpha_{i,t}$ instead of $\theta_{i,t}$, $\theta_{i,t}$ can be determined by Equation~\eqref{eqn-3.10}, $i=0,1,2$. 

Here, we re-experiment the complete model sets from \cite{B:13}, three linear stationary auto-regressive (AR) models with different unconditional means (UM), i.e.,
\begin{align}\notag\label{eqn-4.1}
    \mathcal{M}_{1}:&\quad\by_{1,t}=0.1+0.6\,\by_{1,t-1}+\varepsilon_{1,t},\\
    \mathcal{M}_{2}:&\quad\by_{2,t}=0.3+0.2\,\by_{2,t-2}+\varepsilon_{2,t},\\\notag
    \mathcal{M}_{3}:&\quad\by_{3,t}=0.5+0.1\,\by_{3,t-1}+\varepsilon_{3,t},
\end{align}
with $\varepsilon_{i,t}\overset{i.i.d.}\sim\mathcal{N}(0,\sigma^{2})$, $t=1,\cdots,T$, and assume $\by_{i,0}=0.25$ and $\sigma=0.05$, $i=1,2,3$. The first model $\mathcal{M}_{1}$ is the true data generating process, and the other two are biased predictors.

We realize that the initialization of $\balpha_{0}=(\alpha_{0,0},\alpha_{1,0},\alpha_{2,0})$ significantly affects model performance in the DTVW. To identify optimal parameters, we set $\alpha_{0,0}=0$, i.e. $\theta_{0,0}=0$ by Equation~\eqref{eqn-3.9}, and employ a grid search over $\alpha_{1,0}$ for the adaptive TVW and $(\alpha_{1,0},\alpha_{2,0})$ for the DTVW. We search within the region $(\alpha_{1,0},\alpha_{2,0})\in[-10,10]^2$ with grid size $0.5$, where this range is chosen so that the resulting parameters $(\theta_{1,0},\theta_{2,0})\in[-1,1]^2$. At each grid point, we evaluate performance using CRPS, where $10$ sample paths from the individual predictive distribution is used.

\begin{figure}[htbp]
\centering

\begin{subfigure}[t]{0.49\textwidth}
    \vspace{0pt}
    \centering
    \includegraphics[width=\linewidth]{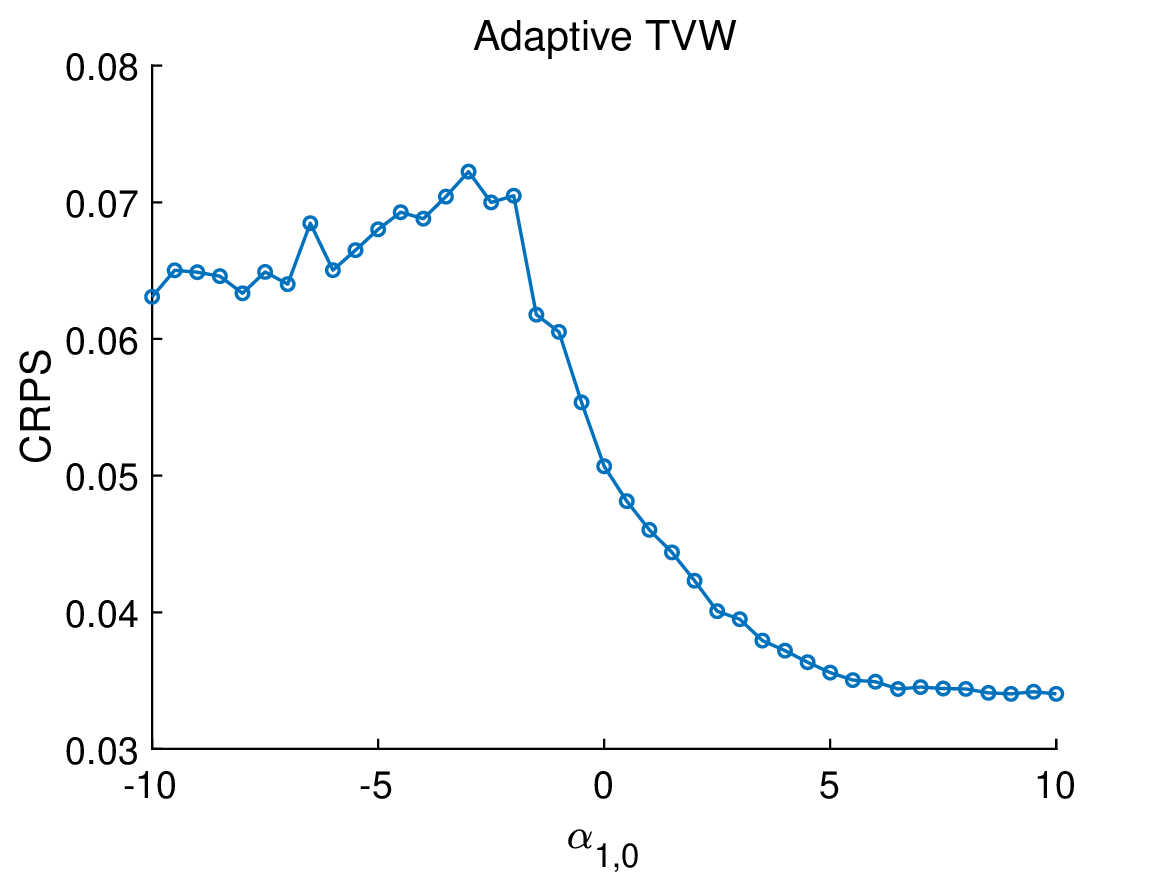}
    \label{fig:simugrid1}
\end{subfigure}
\hfill
\begin{subfigure}[t]{0.49\textwidth}
    \vspace{0pt}
    \centering
    \includegraphics[width=\linewidth]{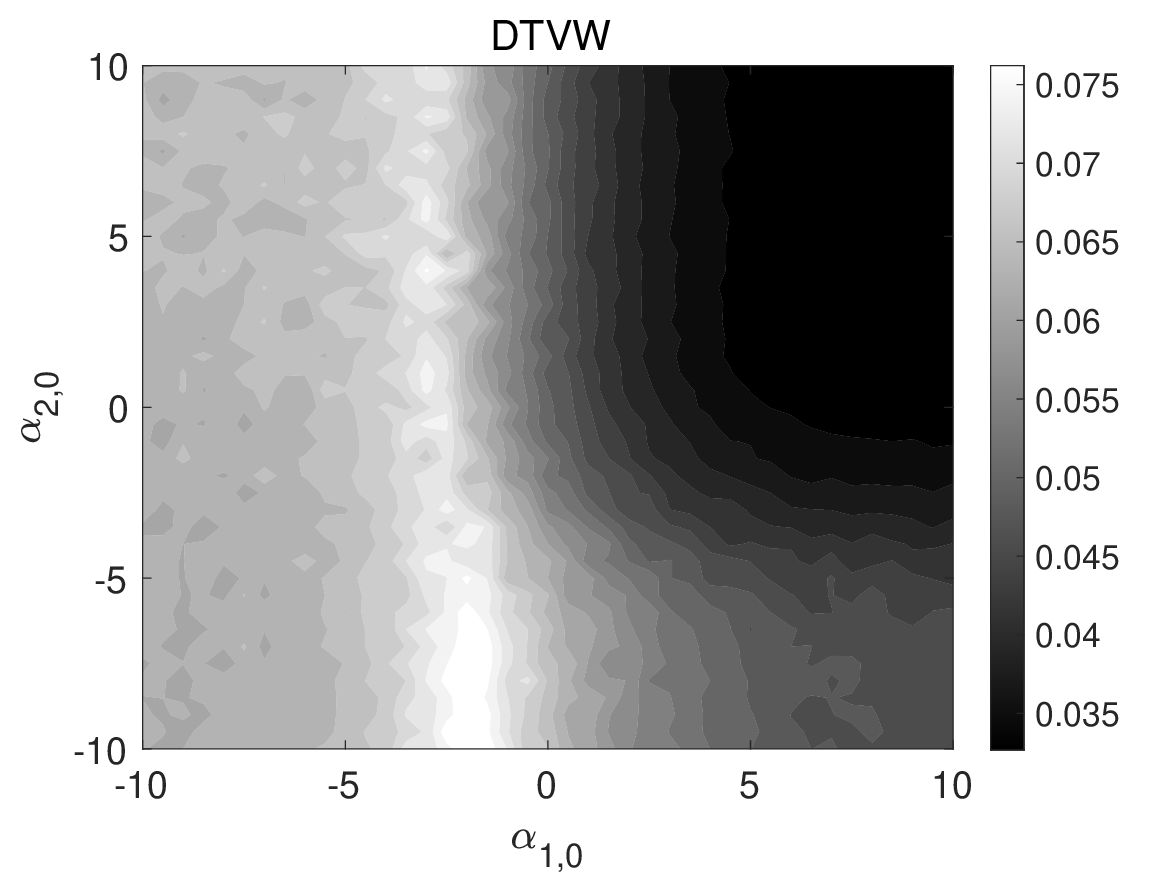}
    \label{fig:simugrid2}
\end{subfigure}
 \vspace{-20pt}
\caption{The CRPS based initialization of $\balpha_{0}$ at each grid point, with $\alpha_{1,0}$ in adaptive TVW and $(\alpha_{1,0},\alpha_{2,0})$ in DTVW.}
\label{fig:crps_combined}
\end{figure}

\begin{figure}[htbp]
\setlength{\abovecaptionskip}{0pt}
\begin{center}
\includegraphics[width=5in]{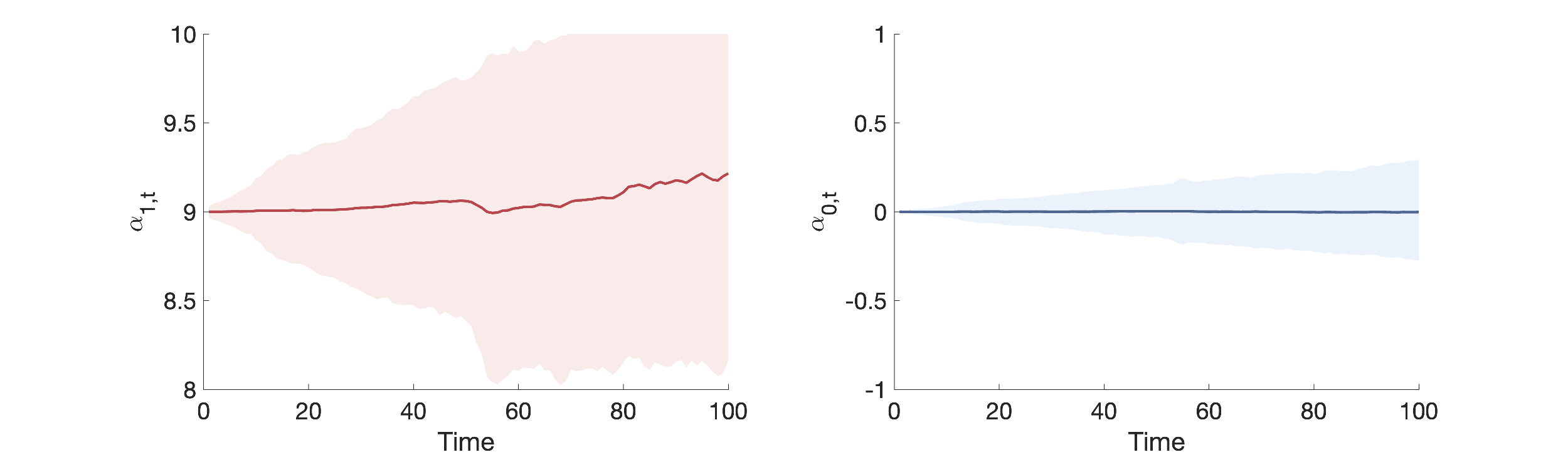}
\end{center}
\small
\caption{The posterior predictive means of the parameters $\alpha_{1,t}$ and $\alpha_{0,t}$ in the adaptive TVW with the initial value $\alpha_{1,0}=9$ (i.e. $\theta_{1,0}=0.9998$) and $\alpha_{0,0}=0$. The shaded area show $95\%$ confidence intervals. 
\label{fig:sec4.1coefficient}}
\end{figure}

In Figure \ref{fig:crps_combined}, we plot the CRPS versus the initialization of $\alpha_{1,0}$ or $(\alpha_{1,0},\alpha_{2,0})$ at the grid point. The left panel shows the CRPS decreases dramatically as $\alpha_{1,0}$ increasing from $0$ to $10$. It suggests that the minimal CRPS is achieved at $\alpha_{1,0}\approx9$, i.e. $\theta_{1,0}\approx0.9998$, which is extremely close to $1$. $\alpha_{1,0}$ is the parameter in front of $\bx_{t-1}$ in the TVW. With this initial parameters, i.e. $\theta_{0,0}=0$, $\theta_{1,0}\approx0.9998$, we apply the PF introduced in Section \ref{sec-3.2.3} to estimate the combination weights for the three models Equation~\eqref{eqn-4.1}, along with their associated confidence intervals conditioned on $1000$ sample paths from the
predictive distribution. Figure \ref{fig:sec4.1coefficient} displays how the parameters $\alpha_{0,t}$ and $\alpha_{1,t}$ change along the time in the adaptive TVW. We find that $\alpha_{0,t}$ remains close to $0$ and $\alpha_{1,t}$ stays around $9$, equivalently $\theta_{1,t}$ keeps around $1$, all the time throughout, except gradually larger confidence intervals. This verifies numerically that our DTVW by setting $\alpha_{2,t}\equiv\theta_{2,t}\equiv0$ and properly chosen the initialization of $(\alpha_{0,0},\alpha_{1,0})$, can recover the TVW in \cite{B:13}. This adaptive TVW gives a data-driven explanation to why $\theta_{0,0}=0$, $\theta_{1,0}=1$ is chosen in TVW. However, the adaptive TVW has fewer free parameters than our DTVW to be tuned to achieve possible better performance. 

\setlength{\textfloatsep}{8pt}

\setlength{\textfloatsep}{8pt}
\begin{figure}[htbp]
\setlength{\abovecaptionskip}{0pt}
\begin{center}
\includegraphics[width=4.5in]{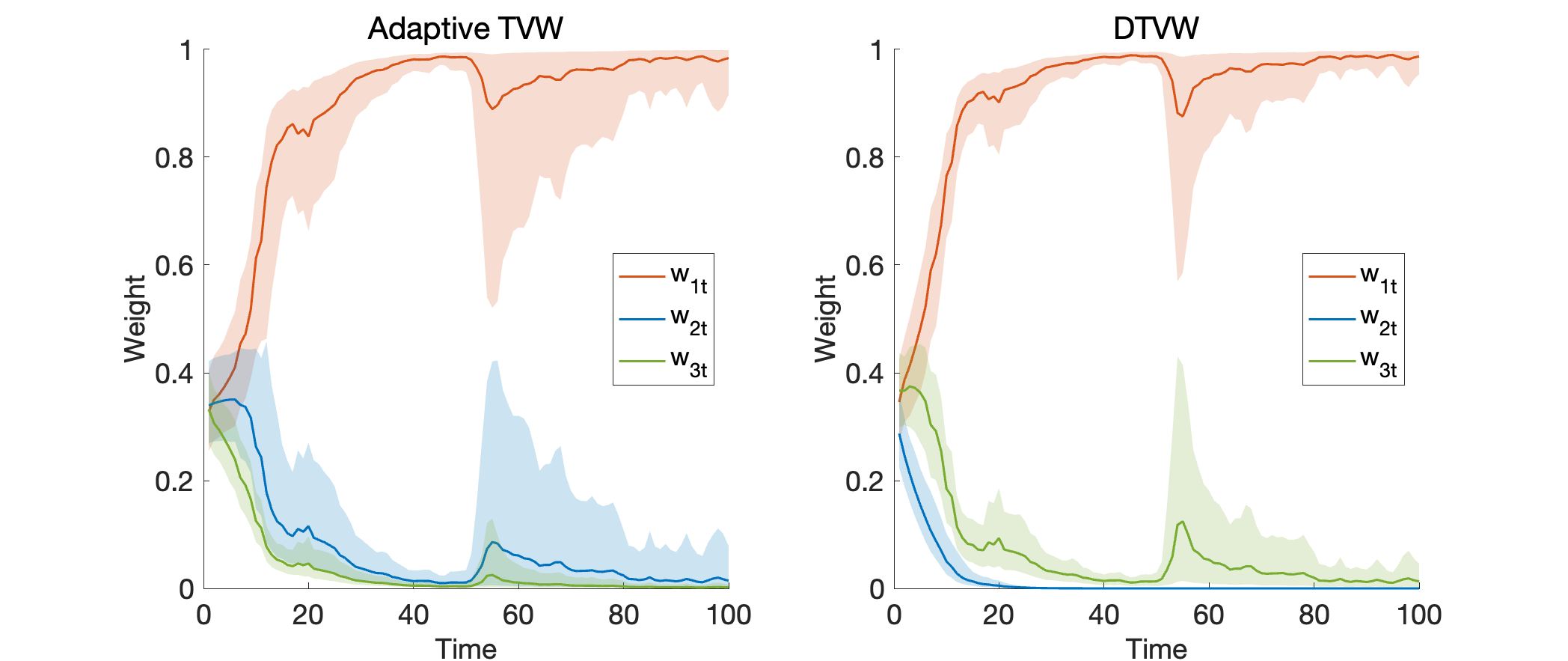}
\end{center}
\small
\caption{The posterior mean of the combination weights and their $95\%$ confidence regions for models $\mathcal{M}_1$-$\mathcal{M}_3$. 
\label{fig:simuwei1}}
\end{figure}

Back to the right panel of Figure \ref{fig:crps_combined}, we plot the CRPS contours with respect to grid points on $(\alpha_{1,0},\alpha_{2,0})\in[-10,10]^2$ of grid size $0.5$. The CRPS is quantized with size $0.005$. The darker shade indicates the
lower CRPS. But the actual CRPS within the same quantized level still vary from point to point. The right panel of Figure \ref{fig:crps_combined} shows that the minimal CRPS is probably achieved within the region  $(\alpha_{1,0},\alpha_{2,0})\in[5,10]\times[0,10]$. By taking a closer look into the actual value of the CRPS, the initial value for our DTVW with the lowest CRPS is $(\alpha_{1,0},\alpha_{2,0})=(10,8.5)$.

As shown in Figure \ref{fig:simuwei1}, after some iterations the weight for the true model $\mathcal{M}_{1}$ converges to $1$ while the weights for the other models converge to $0$ in both the adaptive TVW and DTVW. In Figure \ref{fig:simuwei1}, the weight $\bw_{1,t}$ in DTVW increases to $1$ more rapidly than the adaptive TVW from $t=0$ to $20$. This demonstrates that the DTVW identifies the true model faster than the adaptive TVW. Meanwhile, it yields narrower confidence intervals for the model weights than the adaptive TVW, which reduces the uncertainty in the combination forecast. It suggests that the DTVW has more superior capability of identifying the true model in the simple complete model sets with properly chosen initialization of the parameters.

\setlength{\tabcolsep}{5pt}
\begin{table}[h]
    \centering
    \caption{Forecast accuracy for the biased complete model set Equation~\eqref{eqn-4.1}.}
    \resizebox{\textwidth}{!}{%
        \begin{minipage}{1.2\textwidth} 
            \centering
            \begin{tabular}{lccc@{\hspace{20pt}}ccc}
                 \hline
        &$\mathcal{M}_1$ & $\mathcal{M}_2$ & $\mathcal{M}_3$ & TVW& adaptive TVW& DTVW \\
        \hline
        RMSFE & 0.060&	0.145&	0.318&	0.064&0.064&	{\bf 0.063} \\
        LS & -1.366&	2.858&	32.951&	-0.788&	-0.807& {\bf-0.952}  \\
        CRPS & 0.034&	0.107&	0.285&	0.037&	0.037& {\bf 0.036}   \\
        \hline
            \end{tabular}
            
        \end{minipage}
        }
    \label{tab:Complete_accuracy}
\end{table}

To quantify the performances, we list the RMSFE, LS and CRPS of each models $\mathcal{M}_1$-$\mathcal{M}_3$ and the three combination forecast: TVW, the adaptive TVW and DTVW in Table \ref{tab:Complete_accuracy}. The error of $\mathcal{M}_1$ is from the sampling variance, since it is the true model, while the other two models $\mathcal{M}_2$ and $\mathcal{M}_3$ are mainly due to the model misspecification. It is no wonder that the misspecified models bear much larger errors in any metric. Naturally, no combination forecast can beat the errors of $\mathcal{M}_1$. Yet, all of them yield errors close to it. The adaptive TVW only slightly improve the TVW in LS, while the DTVW achieves a more visible gain in forecast accuracy compared to the other two under all three metrics, suggesting that incorporating diversity plays a role in improving predictive performance.

\setlength{\textfloatsep}{8pt}

\subsubsection{Complex incomplete model sets}\label{sec-4.2.2}

In this subsection, we shall focus on evaluating the predictive performance of the DTVW under a complex nonlinear data-generating process (DGP), with the true model being excluded from the model sets. The true model features strong nonlinearity,  
\begin{equation}\label{eqn-4.2}
\by_t = 0.5\,\by_{t-1} + \dfrac{25 \,\by_{t-1}}{1 + \by_{t-1}^2} + 8\cos(1.2(t-1)) + \varepsilon_{0,t}, 
\end{equation}
which has been used to evaluate nonlinear filters’ performance~\citep{EBPS:15,PMZ:22}.
The model set is constructed as follows:
\begin{align}\notag\label{eqn-4.3}
    \mathcal{M}_{1}:  \quad     \by_{1,t} &= 0.5\, \by_{1,t-1} + \dfrac{15 \,\by_{1,t-1}}{1 + \by_{1,t-1}^2} + 8\cos(1.2(t-1)) + \varepsilon_{1,t}, \\\notag
    \mathcal{M}_{2}:  \quad     \by_{2,t} &= 0.5\, \by_{2,t-1} + \dfrac{25 \,\by_{2,t-2}}{1 + \by_{2,t-2}^2} + 8\cos(1.2(t-1)) + \varepsilon_{2,t}, \\
    \mathcal{M}_{3}:  \quad      \by_{3,t} &= 0.5\, \by_{3,t-1} + \dfrac{5 \,\by_{3,t-1}}{1 + \lvert \by_{3,t-1} \rvert} + 8\cos(1.2(t-1)) + \varepsilon_{3,t}, \\\notag
    \mathcal{M}_{4}:   \quad     \by_{4,t} &= 0.5 \,\by_{4,t-1} + 0.5 \,\by_{4,t-2} + 9\cos(1.2(t-1)) + \varepsilon_{4,t}, \\\notag
    \mathcal{M}_{5}: \quad      \by_{5,t} &= \by_{5,t-1} + 8\cos(1.2(t-1)) + \varepsilon_{5,t}, \\\notag
    \mathcal{M}_{6}:   \quad     \by_{6,t} &=  \dfrac{30 \by_{6,t-1}}{1 + \by_{6,t-1}^2} + 8\cos(1.2(t-1)) +\varepsilon_{6,t},
\end{align}
where $\varepsilon_{it}\overset{i.i.d.}\sim\N(0,\sigma^{2})$, $t=1,\cdots,T$, and assume $\by_{i,0}=0.1$ and $\sigma=0.5$, $i=1,\cdots,6$.

\begin{figure}[!ht]
\centering
\includegraphics[width=0.47\linewidth]{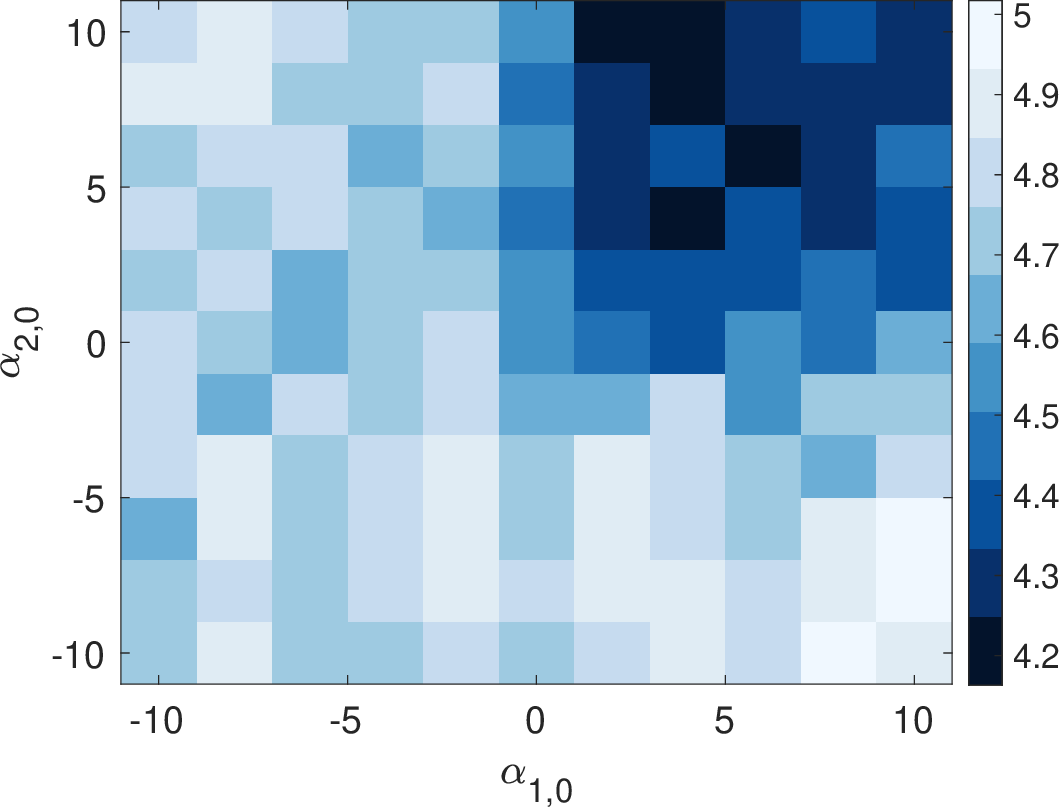}
\includegraphics[width=0.47\linewidth]{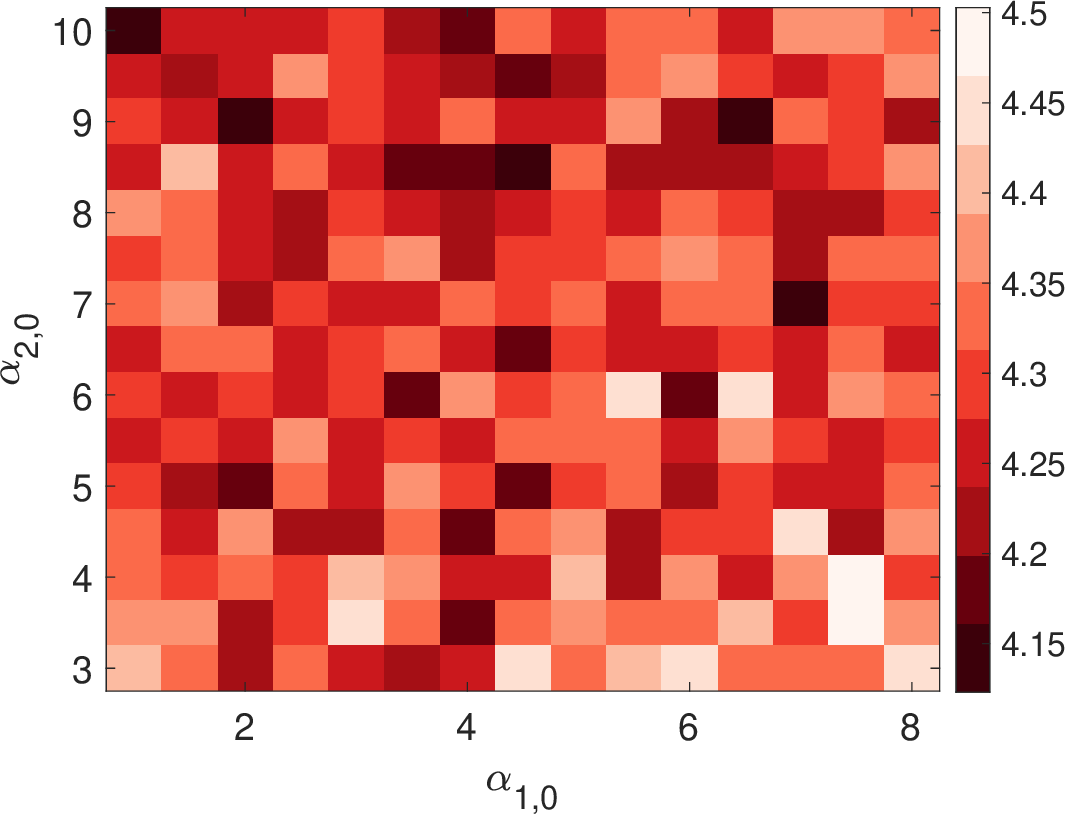}
\caption{Two-stage grid search for the initial values $(\alpha_{1,0},\alpha_{2,0})$ in DTVW for the incomplete model sets Equation~\eqref{eqn-4.3}.}
\label{fig:crps_heat4.2}
\end{figure}

As in Section \ref{sec-4.2.1}, we again employ a grid search to determine the initial values for $\alpha_{1,0}$ and $\alpha_{2,0}$ in our DTVW. Not like the AR model before, due to the stronger nonlinearity of the DGP, a grid search with a fine size would be computationally intensive. To improve efficiency, we adopt a two-stage grid search strategy illustrated in Figure \ref{fig:crps_heat4.2}. That is, a coarse grid search over the region $[-10,10]^2$ with step size $2$ is performed first to locate the darkest region $[1,8]\times[3,10]$ in the left panel. Then a second round grid search with finer step size $0.5$ is performed in this region in the right panel. Compared the heatmap in the right panel with that in Figure \ref{fig:crps_combined}, there is no monotonicity can be observed in the changes of the CRPS. There are several spots comparatively dark. Finally, we set the lowest CRPS's spot $(\alpha_{1,0},\alpha_{2,0})=(7,7)$ as the initial value in our DTVW below.

\setlength{\tabcolsep}{5pt}
\begin{table}[h!]
    \centering
    \small
    \caption{Forecast accuracy for the complex nonlinear incomplete model set Equation~\eqref{eqn-4.3}.}
    \resizebox{\textwidth}{!}{%
        \begin{minipage}{1.1\textwidth} 
            \centering
            \begin{tabular}{lcccccc@{\hspace{20pt}}cc}
                 \hline
        &$\mathcal{M}_1$ & $\mathcal{M}_2$ & $\mathcal{M}_3$ & $\mathcal{M}_4$ & $\mathcal{M}_5$ & $\mathcal{M}_6$ &TVW& DTVW \\
        \hline
        RMSFE &8.081&	8.472&	8.871&	9.789&	9.711&	10.112&	7.742&	{\bf7.554} \\
        LS & 10.658& 	3.992&	8.335&	8.995&	6.134&	4.316&	2.980&	{\bf2.533} \\
        CRPS & 5.052&	4.901&	5.642&	6.584&	6.010&	6.689&	4.455&	{\bf3.984}  \\
        \hline
            \end{tabular}            
        \end{minipage}
        }
    \label{tab:nonliner_accuracy_4.2}
\end{table}

Table \ref{tab:nonliner_accuracy_4.2} reports the forecast errors of the individual models $\mathcal{M}_1$-$\mathcal{M}_6$ and two combination forecast: TVW and DTVW, for the complex nonlinear incomplete model sets Equation~\eqref{eqn-4.3}. As shown in this table, the individual models perform much worse than the combination ones, and our DTVW consistently outperforms the TVW in all three evaluation metrics: RMSFE, LS and CRPS are reduced by $2.43\%$, $15.01\%$ and $10.57\%$, respectively. Back to the simple complete model set Equation~\eqref{eqn-4.1}, Table \ref{tab:Complete_accuracy} shows that the DTVW still delivers modest gains over TVW in RMSFE and CRPS by $1.56\%$ and $2.70\%$, respectively. The relative pronounced improvement in LS here is by $20.83\%$, which is more marked than $15.01\%$ in Table \ref{tab:Complete_accuracy}. We believe that this phenomenon is primarily due to the much lower baseline error in the simpler model sets. It may indicate that our DTVW is especially effective in complex or misspecified forecast environments, where its diversity-aware weighting mechanism enables better exploitation of information across models.

\begin{figure}[h!]
\begin{center}
\includegraphics[width=5in]{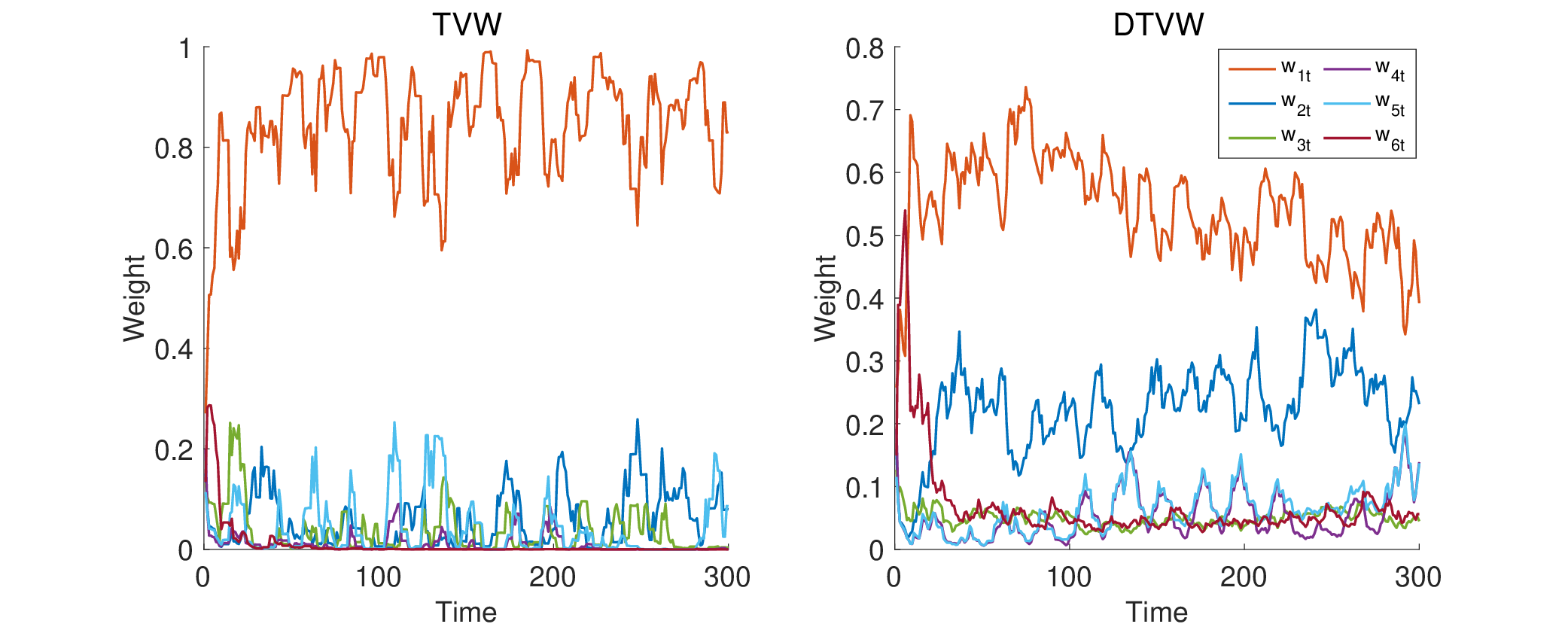}
\end{center}
\caption{The posterior mean of the combination weights for models $\mathcal{M}_1$-$\mathcal{M}_6$ in the complex incomplete model sets Equation~\eqref{eqn-4.3}.}
\label{fig:weight4.2}
\end{figure} 

To understand why the DTVW outperforms the TVW in complex incomplete model sets, we plot the estimated combination weights in Figure \ref{fig:weight4.2}, with the TVW's weights in the left panel, and the DTVW's in the right one. The TVW puts most of the weight on model $\mathcal{M}_1$ and assigns very low weights to the others. This may be explained by the fact that the RMSFE of $\mathcal{M}_1$ is the best among all six models, displayed in Table \ref{tab:nonliner_accuracy_4.2}. But when it comes to metrics related to the density estimation, $\mathcal{M}_2$ is the best in both LS and CRPS. Thus, the low weight assigned to $\mathcal{M}_2$ in the TVW seems to be inappropriate. On the contrary, the DTVW captures this by assigning a relatively large weight to $\mathcal{M}_2$, second only to $\mathcal{M}_1$. This improves the overall forecast performance.

By assigning non-negligible weights to relative better performance models, the DTVW achieves a more balanced and robust forecast. This adaptive weighting mechanism reduces the risk of over-believing a single dimension of performance and enhances generalization across different evaluation criteria. In complex or misspecified environments where the true DGP is not among the candidate models, diversity plays a crucial role. By adopting a “forecast with forecasts” approach, it not only expands the historical information in model combination, but also captures the information in future from individual forecast models, leading to more balanced and adaptive weights. These findings suggest that diversity-aware forecast combination serves as a more general framework than traditional weighting schemes. It not only encompasses conventional approaches as special cases but also demonstrates superior performance in more complex and uncertain forecast environments. 

\section{Empirical applications}\label{sec-5}

This section presents results from real-time out-of-sample forecast exercises, generating both point and density forecasts for empirical datasets.  

\setcounter{equation}{0}

\subsection{The Real Price of Oil}\label{sec-5.1}

\begin{figure}[h!]
\setlength{\abovecaptionskip}{-10pt} 
\begin{center}
\hspace*{-0.03\textwidth}
\includegraphics[width=5in]{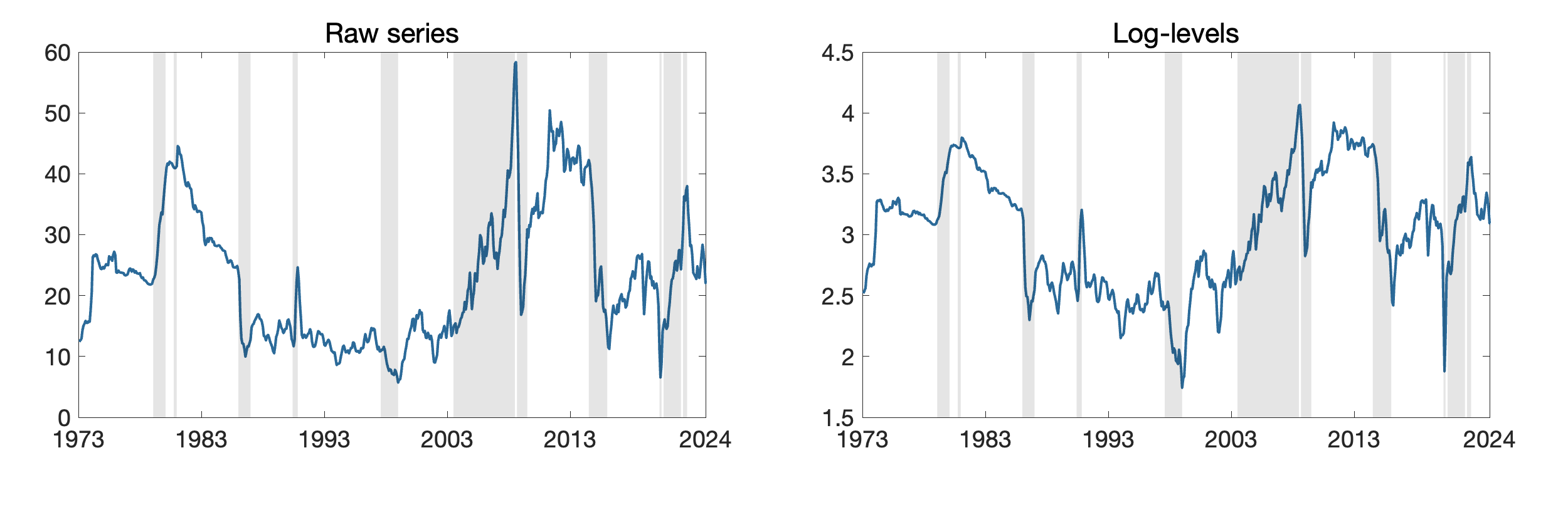}
\end{center}
\captionsetup{width=\linewidth}
\caption{Real monthly IRAC oil price over the period 1973:01-2024:08.}
\label{fig:Oildata}
\end{figure}

In this subsection, we shall forecast the real price of oil in the global market. The monthly real-time data in \citet{KJH:23} and \citet{GVZ:19} are extended to 2024:08, such that the most recent oil price would also be investigated. Specifically, we adjust the U.S. refiners' acquisition cost for imported crude oil (IRAC) for inflation using the seasonally adjusted U.S. Consumer Price Index (CPI). The real oil price and its logarithmic transformation are displayed in Figure \ref{fig:Oildata}. The log transformation stabilizes variance and removes the exponential trend, facilitating clearer identification of percentage-based price shifts and long-term structural changes rather than transient volatility. The shaded regions highlight critical episodes of oil market volatility, including the 1979 oil crisis, the 1980 Iran-Iraq War, OPEC’s 1985 policy shift to increase production, the 1990-1991 Gulf War, the 1997-1998 Asian Financial Crisis, the 2003–2008 surge, the Great Recession collapse, the 2014–2016 price collapse, the COVID-19 demand shock (2020:03-2020:05), OPEC+ supply cuts (2020:06-2021:12), the Russia-Ukraine War (2022:02-2022:06), and renewed OPEC+ restrictions (2024:03-2024:08).

Three patterns are observed and analyzed as follows. First, the log-level series shows persistent shifts in average prices over time, such as the prolonged rise after the 1979 oil crisis and the 2003–2008 surge. These changes suggest that static models may fail to capture long-term trends, supporting the use of time-varying forecast approaches. Second, sudden price shifts, such as the 2014–2016 drop due to oversupply or the 2022 volatility from the Russia-Ukraine War, align closely with political conflicts and economic disruptions, highlighting how real-world events drive structural breaks. Third, after 2018 the real oil price series displays sharper reversals and greater short-term variability. This intensified pattern driven by the collapse of COVID-19 demand, which continued through cuts in OPEC+ production, reflects pressures from changes in energy policy, impacts of pandemics, and ongoing supply constraints. Together, these findings emphasize the need for flexible models, adapt to the dynamics with both gradual trends and sudden shocks. 

Following \cite{KJH:23},
the six state-of-the-art individual oil price forecast models are included to obtain the point forecasts: a no-change forecast (NC), a model related to the changes in the price index of non-oil industrial raw materials (CRB), a model related to West Texas Intermediate oil futures prices (Futures), a model related to the spread between spot prices of gasoline and crude oil (Gasoline), a model related to time-varying parameter of the gasoline and heating oil spreads (TVspread), and an Vector Autoregressive Model (VAR). We refer the interested readers to \cite{KJH:23} for a detailed introduction of these individual models. The initial period 1973:01 to 1991:12 is for estimation in individual models. Starting from 1992:01, the monthly real-time forecasts are yielded until 2024:08. To obtain the density forecasts from the point ones, the Metropolis-within-Gibbs sampling algorithm described in \cite{CH:14}
is used to obtain $1000$ draws from the posterior distribution after discarding the first $5000$ draws as burn-in. Then, the density forecasts from the last five individual models, including CRB, WTI, Gasoline, TVspread and VAR, are combined, whilst the NC model is used as the baseline for performance comparison. 

We applied five combination models: the Equal Weight method, the Bayesian Model Averaging (BMA) and a 24-month rolling window BMA described in Section \ref{sec-2.3}, the Time-Varying Weights (TVW) method and our proposed Diversity-driven Time-Varying Weights (DTVW) method described in Section \ref{sec-3.2.2}. Equal initial weights are assigned to the five individual models, and each set should be $1/5$. $N=1000$ particles are used in the particle filter in TVW and DTVW. In the next subsection, we shall compare the predictive performance of all the combination forecast methods mentioned above.

\begin{figure}[htbp]
\centering
\includegraphics[width=0.45\linewidth]{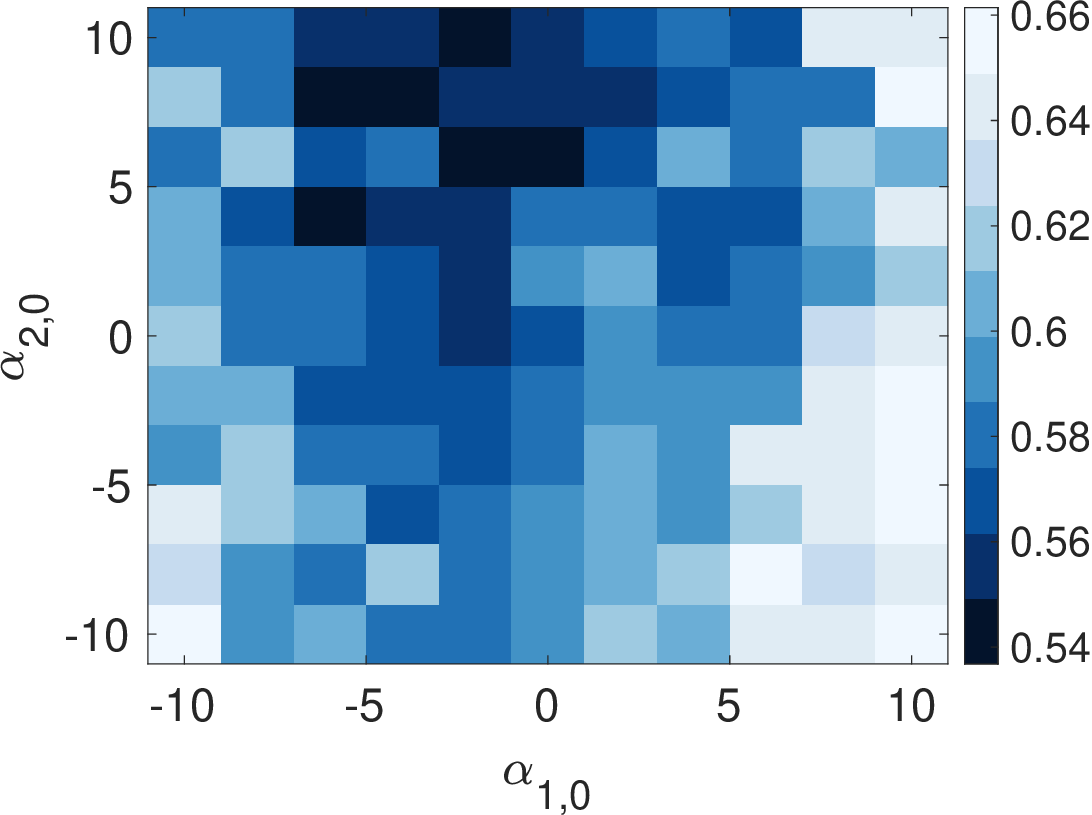}
\includegraphics[width=0.45\linewidth]{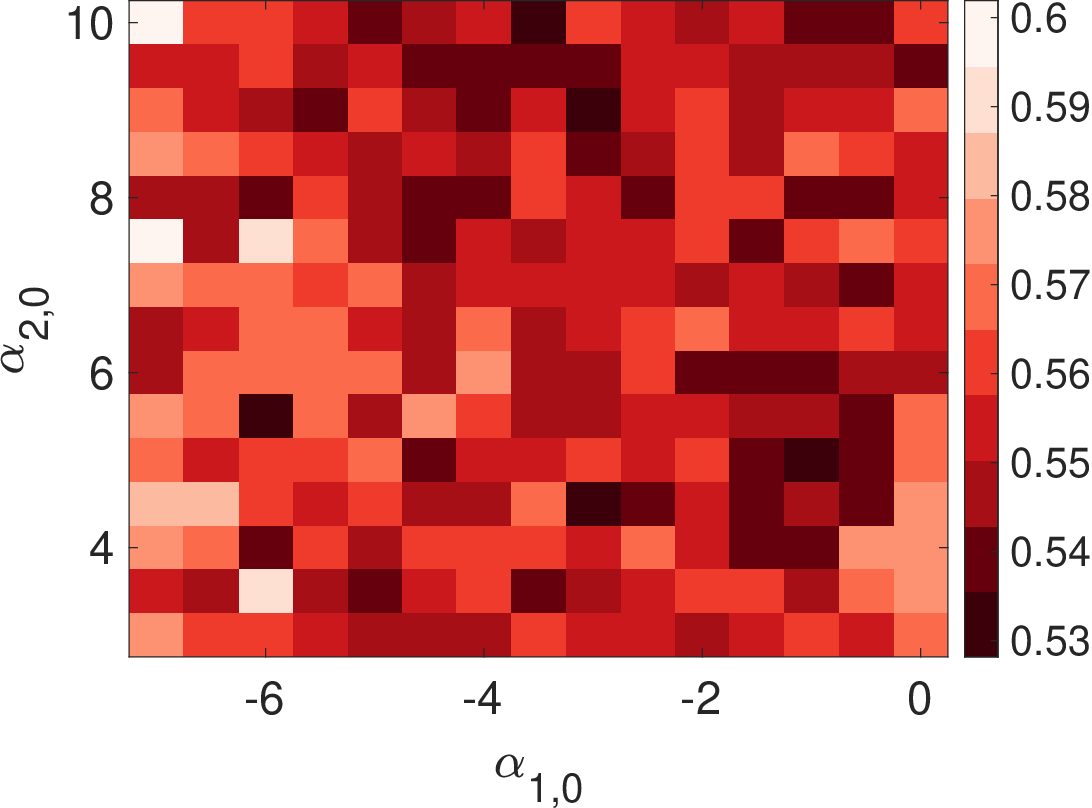}
\caption{Two-stage grid search for the initial values $(\alpha_{1,0},\alpha_{2,0})$ in the DTVW for the oil price.}
\label{fig:crpsoil_heat}
\end{figure}

As illustrated in Section \ref{sec-4.2}, the initialization of the forecast parameters in a step ahead $(\alpha_{1,0},\alpha_{2,0})$ in DTVW is crucial to the performance. In this section, we perform the two-stage grid search as in the complex incomplete model set \eqref{eqn-4.3}; see Section \ref{sec-4.2.2}. The left panel of Figure \ref{fig:crpsoil_heat} illustrates the coarse search in the first stage conducted in the domain $[-10, 10]^2$ with a grid size of $2$. This analysis indicates that the subdomain $[-7, 0] \times [3, 10]$-to be explored with a refined grid size of $0.5$-merits further detailed investigation. The right panel presents the refined search focused on this specific region. The parameters with the lowest CRPS are located at $(\alpha_{1,0},\alpha_{2,0})=(-3,4.5)$. This heatmap, similar as that in Figure \ref{fig:crps_heat4.2}, exhibits no monotonicity. This may imply that the oil price is in a more complex and misspecified forecast environment. A similar two-stage grid search procedure can also be applied for multi-step ahead forecasts.

Forecast performance is evaluated at multiple horizons (1-step, 3-step and 6-step ahead) to assess the predictive precision of all individual and combination forecast models. This empirical study aims to demonstrate two key points: first, our DTVW is capable of achieving superior performance in forecast commodity prices, which is difficult as all we known; second, our DTVW offers consistent improvements in both 1-step and multi-step predictions.

\subsubsection{Forecast accuracy analysis: 1-step and multi-step predictions}

First, we focus on the forecast accuracy in three metrics defined in Section \ref{sec-4.1}: RMSFE, LS and CRPS. The lower values indicate better performances. In addition, DM tests~\citep{DM:95} are performed to assess the statistical significance of differences in predictive accuracy. 

\setlength{\tabcolsep}{3pt}
\begin{table}[h!]
    \centering
    \small
    \caption{Forecast accuracy for the oil price}
    \resizebox{\textwidth}{!}{%
        \begin{minipage}{1.3\textwidth} 
            \centering
            \begin{tabular}{ccccccc@{\hspace{15pt}}ccccc}
             \hline
             \multicolumn{12}{c}{\textbf{RMSFE}} \\
                 \hline
        \textbf{Horizon} &NC& CRB & Futures & Spread & TVspread & VAR &Equal& BMA & BMA\_roll & TVW& DTVW \\
        \hline
        1 & 2.331& 	2.237 &	2.291 &		2.331 &	2.366 &	2.307&	2.235$^{**}$& 2.235$^{**}$&2.145$^{**}$ &	1.507$^{**}$ & \textbf{1.337}$^{**}$ \\
        3 & 4.530&	4.264$^{*}$&	4.447&	4.601 	&4.657$^{*}$&	4.730&	4.371 &4.734&	4.224$^{**}$ &3.404$^{**}$ & \textbf{3.032}$^{**}$ \\
        6 & 6.569&	6.778&	6.367$^{*}$&	6.656 	&6.739&	6.888&	6.413$^{**}$& 6.412$^{**}$&	6.271$^{**}$ &5.218$^{**}$ & \textbf{4.682}$^{**}$ \\
        \hline 
        \multicolumn{12}{c}{\textbf{LS}} \\
                 \hline
        \textbf{Horizon} &NC& CRB & Futures & Spread & TVspread & VAR &Equal& BMA & BMA\_roll & TVW& DTVW \\
        \hline
        1 & 1.931 &1.930&	1.936$^{*}$&	1.941 	&1.948 &1.982$^{**}$ &	2.988$^{**}$ &	2.990$^{**}$ &2.150 	&0.849$^{**}$ & \textbf{0.643}$^{**}$ \\
        3 & 2.451&	2.418&	2.431&	2.452 	&2.442&	2.607$^{**}$&	3.160$^{**}$&3.128$^{**}$&	2.432$^{**}$ 	&1.101$^{**}$ & \textbf{0.732}$^{**}$ \\
        6 & 2.761&	2.754&	2.732$^{*}$&	2.777 	&2.793&	2.857$^{**}$&	3.332$^{**}$& 3.324$^{**}$&	2.723 	&1.351$^{**}$ & \textbf{0.902}$^{**}$ \\
        \hline 
        \multicolumn{12}{c}{\textbf{CRPS}} \\
                 \hline
        \textbf{Horizon} &NC& CRB & Futures & Spread & TVspread & VAR &Equal & BMA & BMA\_roll & TVW& DTVW \\
        \hline
        1 & 1.131&	1.116&	1.121&	1.135$^{*}$ 	&1.147&	1.164&	1.230$^{**}$& 1.230$^{**}$&	1.135 	&0.598$^{**}$ & \textbf{0.511}$^{**}$ \\
        3 & 2.006&	1.935$^{*}$&	1.962&	2.026 	&2.032$^{*}$&	2.246&	 2.180$^{**}$&2.176$^{**}$&	1.938 	&1.093$^{**}$ & \textbf{0.898}$^{**}$ \\
        6 & 2.857&	2.869&	2.757&	2.908$^{*}$ &2.937&	3.113&	 3.159$^{**}$&3.154$^{**}$&	2.762 	&1.568$^{**}$ & \textbf{1.284}$^{**}$ \\
        \hline 
            \end{tabular}
            \\[3pt]
            \raggedright Note: Bold numbers show the most accurate forecasts. $*$ indicates the result is better than the no-change model at $5\%$ confidence level using Diebold-Mariano test, and $**$ is at $1\%$ level.
        \end{minipage}
    }
    \label{tab:oil_accuracy}
\end{table}

Table \ref{tab:oil_accuracy} reports the forecast accuracy across all horizons in different metrics. Our proposed DTVW consistently achieves the best performance across all horizons, with the lowest error values in bold. In terms of RMSFE, the DTVW reduces the point forecast errors by $11.3\%$, $10.9\%$ and $10.3\%$ at the 1-step, 3-step, and 6-step horizons, respectively. As for the probabilistic forecast, the LS's gains are even more pronounced with $24.3\%$, $33.5\%$ and $33.2\%$, respectively. Similarly, the CRPS is reduced by $14.5\%$, $17.8\%$ and $18.1\%$ across the three horizons. These results highlight that DTVW delivers greater improvements over TVW in density forecasts across all multistep forecasts than those in point forecasts. It is worthy of notice that in probabilistic forecast the gains tend to increase in multi-step predictions, which underscoring its robustness in handling extended prediction steps. Moreover, the DM tests show that the DTVW's improvements are credibly different from zero at the $99\%$ confidence intervals.

As for the other forecast combinations, consistent with the findings in \cite{KJH:23}, the BMA performs substantially worse than the NC in probabilistic forecasts across all horizons and yields results that are close to the Equal. Additionally, the rolling-window version of BMA (BMA\_roll) generally outperforms the standard BMA. This may reflect that BMA\_roll offers greater flexibility in adapting to potential shifts in the commodity market. Instead of using all historical information in the BMA, it only utilizes those in the 24-month window and responds more promptly to recent changes.

\begin{figure}[ht!]
\setlength{\abovecaptionskip}{-10pt} 
\hspace*{-0.06\textwidth} 
\begin{center}
\includegraphics[width=5in]{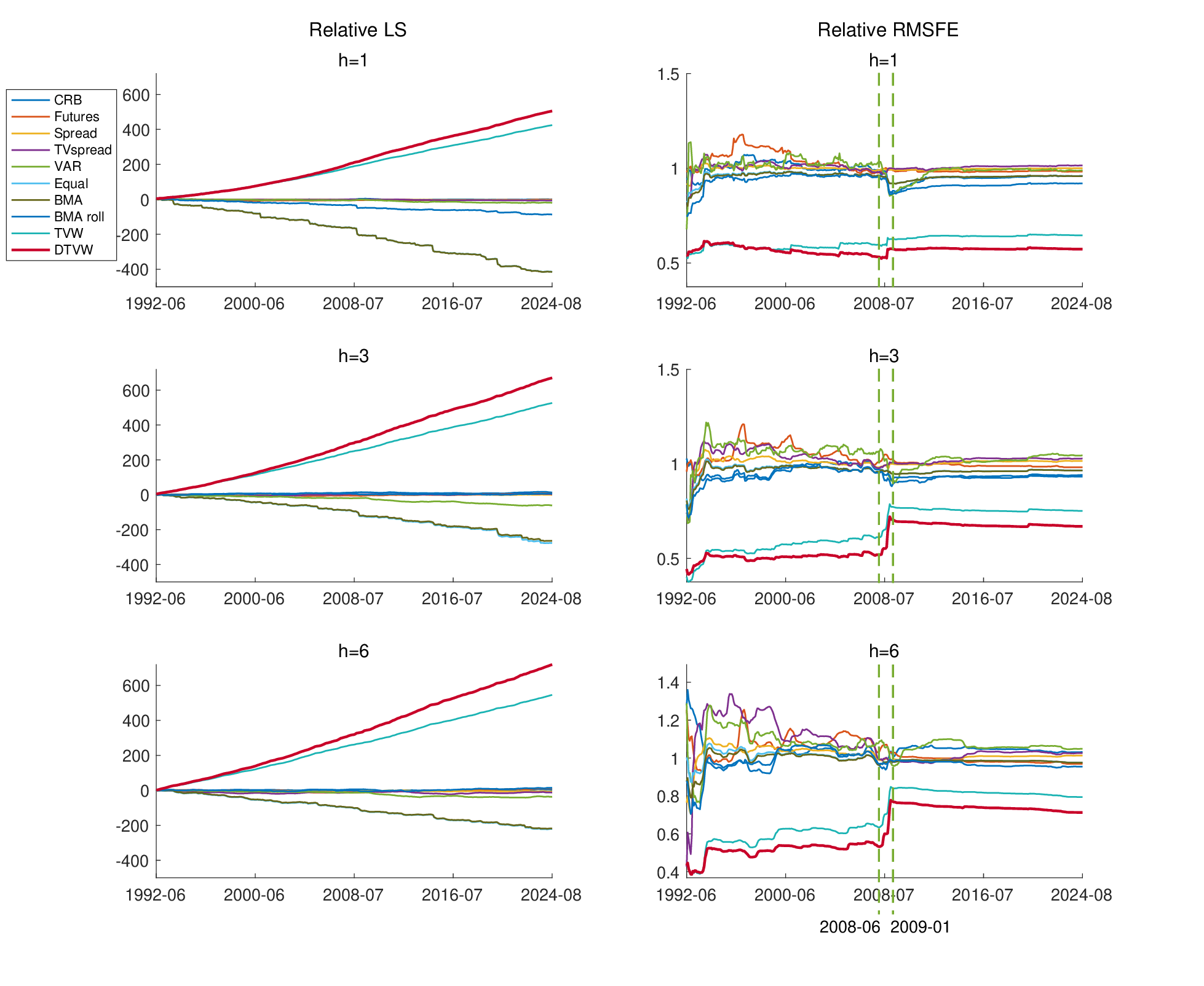}
\end{center}
\small
\caption{Error comparisons among different models in two metrics over the forecast evaluation period: 1992:06–2024:08. The left columns are time series of negative cumulative LS differences with respect to the NC model. The right columns are those of the RMSFE ratios with respect to the NC model. The three rows correspond to the different forecast horizons of 1-month, 3-month and 6-month ahead, respectively.
\label{fig:oil_comparison}}
\end{figure}

Second, we examine how the forecast accuracy varies over time across all models: both individual and combination forecast models. Figure \ref{fig:oil_comparison} shows that the DTVW exhibits superior performance at each time instant across all horizons in both relative LS and relative RMSFE. The left panel displays the difference between the negative cumulative LS of each model and that of the NC model. The larger value indicates better performance. Both TVW and DTVW exhibit obvious upward trend of the curves in all three horizons. This aligns with the intuition that longer term dynamics may be more amenable to structured combination approaches, which leverage diverse information sources. Typically, our DTVW amplifies this advantage: the gap between DTVW and other models widens with $h$, suggesting it is adept at capturing persistent and long horizon patterns. This may be explained by its ability to incorporate future predictive information as a forward-looking signal, aligning the weight estimation with the time-varying information structure of the forecast horizon and allowing greater precision. The right panel shows the RMSFE of each model divided by that of the NC model. The lower value implies better accuracy. A sudden jump occurs around $2008$-$2009$ in all three horizons, a year marked by a sharp collapse in oil prices. The upward jumps are more evidenced in both TVW and DTVW for $h=3$ and $h=6$ than for $h=1$. A closer examination of the top figure for $h = 1$ reveals that the most abrupt drops in the relative RMSFE between 2008:06 and 2009:01 occur in two individual models: the CRB and VAR. This, in turn, enhances the performance of the BMA\_roll, BMA, and Equal. In contrast, the performance of the TVW and DTVW models deteriorates slightly in terms of this point forecast metric. For $h=3$ and $h=6$, such sudden drops in individual models are less pronounced, and the same applies to BMA\_roll, BMA, and Equal. Meanwhile, the performance of both TVW and DTVW models deteriorates even further. We attribute this phenomenon to the following mechanisms: For 1-step ahead forecasts during extreme market shocks, which utilize recent short-term data, certain individual forecast models are capable of capturing structural changes. In contrast, multi-step ahead forecasts accumulate greater uncertainty. Extreme events such as the 2009 oil price spike further disrupt the long-term trends that underpin these multi-step forecasts, while abrupt and significant market fluctuations invalidate the historical patterns on which these models depend. Consequently, the improved performance of specific individual models enables some combined forecast approaches to yield better point predictions, though others fail to benefit from such improvements. Nevertheless, abrupt changes, such as that around 2009, fail to be identified in the density forecast metric-namely, the relative LS-across all three horizons, as shown in the left column.

\begin{figure}[h!]
\begin{center}
\hspace*{-0.08\textwidth} 
\includegraphics[width=5.5in]{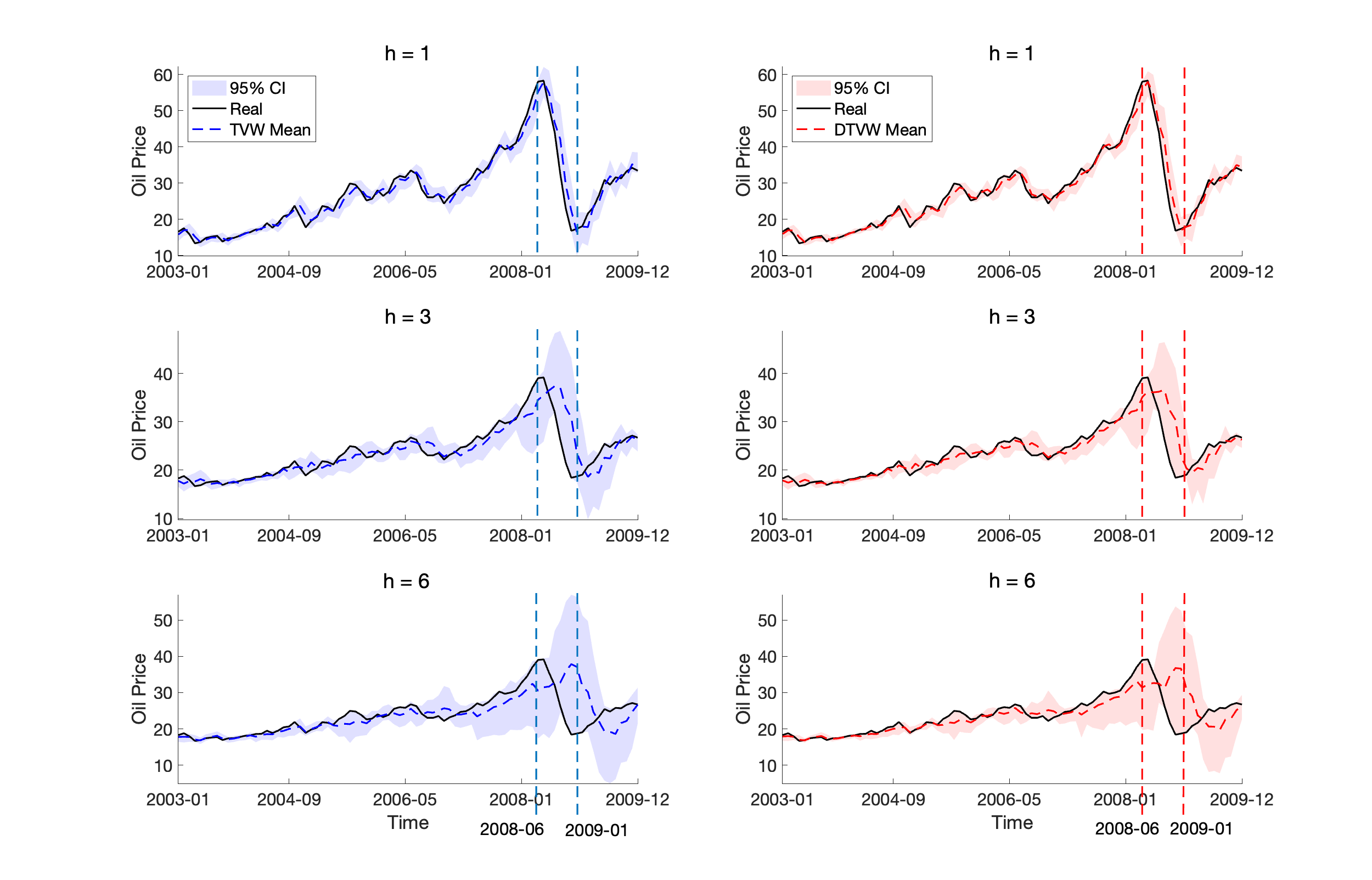}
\end{center}
\caption{The real oil price (solid black lines), the 1-step ahead forecasts' means (dashed lines) by TVW (left column) and DTVW (right column) and their $95\%$ confidence intervals (shaded areas) from 2003 to 2009. 
\label{fig:oil_real}}
\end{figure}

To understand why the density forecast metric is not sensitive in all horizons, we plot the TVW and DTVW forecasts during the oil boom period against the actual oil price in Figure \ref{fig:oil_real}, as well as the 95\% confidence intervals. Between 2008:06 and 2009:01, there are two structural turning points that correspond to the onset and conclusion of the abrupt price decline. Although the mean forecasts of the TVW and DTVW models deviate further from the true price in multi-step ahead forecasts, their 95\% confidence intervals across all horizons still encompass the true price. This implies that the probability of the true price in these intervals remains high ($\geq0.95$) in all cases, without any abrupt fluctuations. The above analysis clarifies why the point forecast metric is more sensitive to abrupt changes, whereas the density forecast metric is not. Furthermore, across all horizons the 95\% confidence intervals generated by the DTVW model are narrower than those of the TVW, indicating reduced forecast uncertainty. This improvement can be attributed to the adoption of a dynamic prior in the DTVW model that incorporates predictive information of future models, enabling more responsive weight updates and more accurate uncertainty quantification during periods of increased volatility. 

\subsubsection{Model selection capability}

In this subsection, we examine whether our DTVW is capable of distinguishing between well-performing and poorly-performing forecast models by analyzing their time-varying weights, as discussed in Section \ref{sec-4.2.2}.

\begin{figure}
\begin{center}
\hspace*{-0.11\textwidth} 
\includegraphics[width=4.5in]{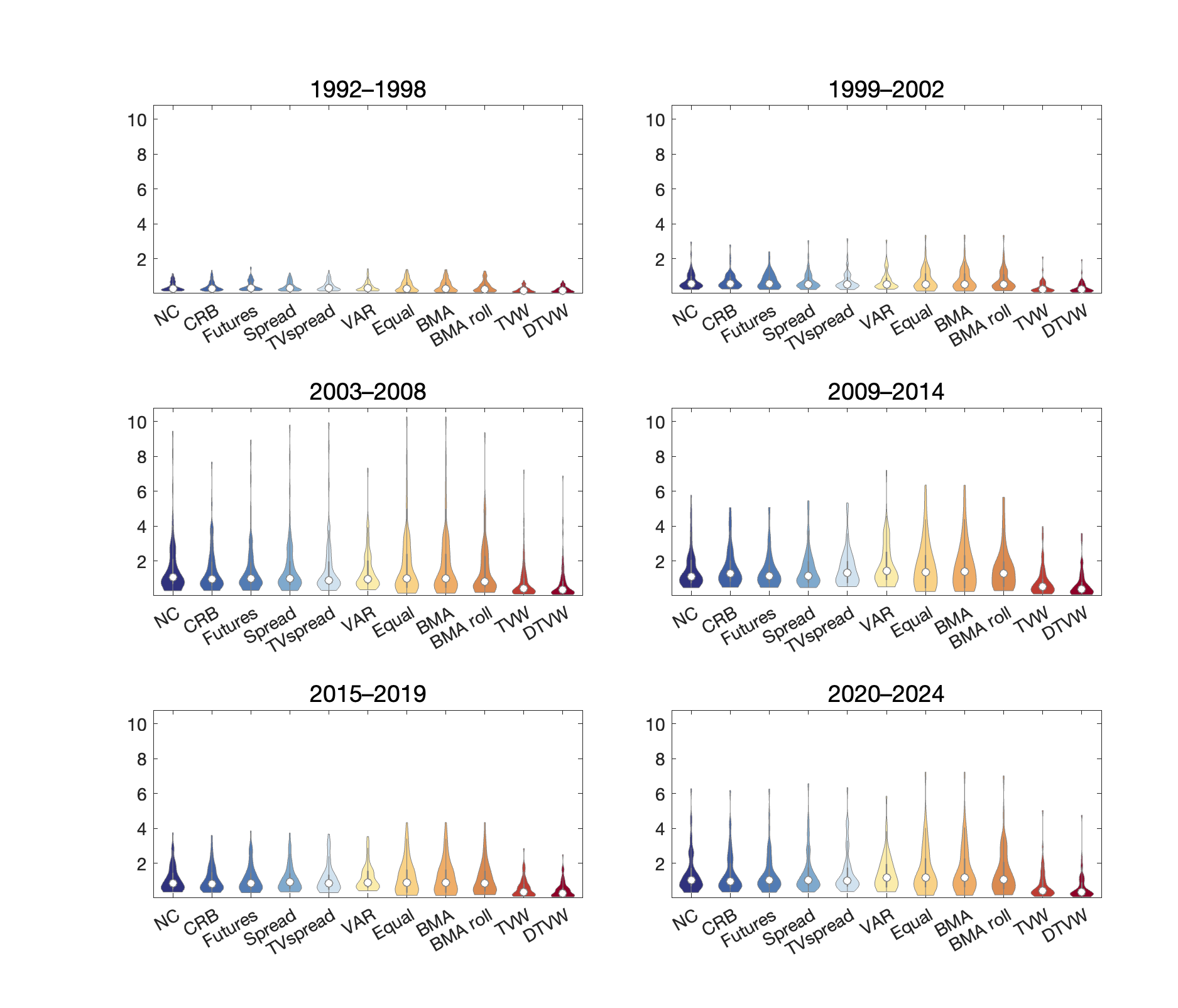}
\end{center}
\caption{The violin plot of different models' CRPS values during six time periods for oil price forecasts. 
\label{fig:oil_violin}}
\end{figure}

To evaluate forecast accuracy under varying market regimes, Figure \ref{fig:oil_violin} displays the CRPS values' distributions across six time periods in succession in a violin plot, revealing changes in the medians and densities of forecast errors. The dot in the middle is the median of the CRPS values. The width of the violin at any point represents the number of observations (or density) around that value; thus wider sections indicate a higher concentration, while narrower sections indicate fewer observations.

The six time periods reflect major structural shifts in the global oil market: post-Cold War stability (1992-1998), early demand recovery (1999-2002), the oil price boom and global financial crisis (2003-2008), post-crisis high price stability (2009-2014), the shale-driven adjustment (2015-2019), and pandemic/geopolitical volatility (2020-2024). The CRPS values during the first two periods 1992-1998 and 1999-2002 have the smallest medians and the narrower shapes, suggesting less variability. In contrast, during the remaining four six time periods, the CRPS values are wider and more dispersed, indicating greater variability. Among these four, the oil boom period (2003–2008), which includes the global financial crisis, exhibits the most wide-spreading and skewed distribution. Nevertheless, among all six time periods, our DTVW yields notably wider and the lowest medians near zero, indicating that the CRPS values are tightly clustered at lower levels. In contrast, other combination forecast models, like Equal, BMA and BMA\_roll, show taller and more dispersed distributions, reflecting larger forecast uncertainty and less consistent performance.

\begin{figure}
\begin{center}
\hspace*{-0.11\textwidth} 
\includegraphics[width=6in]{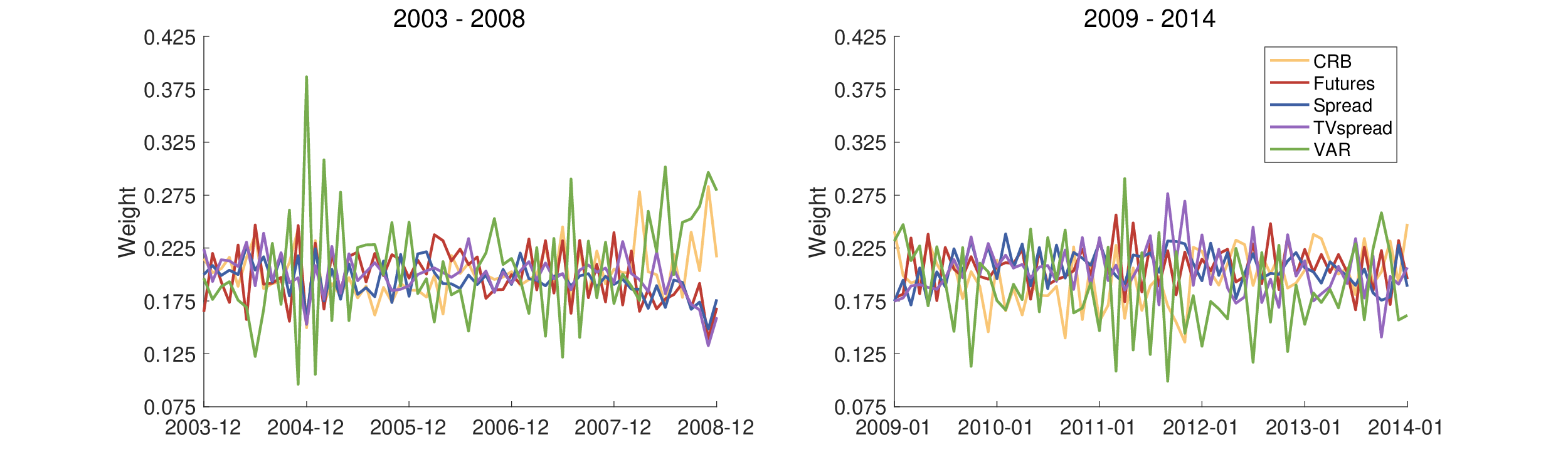}
\end{center}
\caption{The weights' means of different individual models in the DTVW during the oil boom period 2003-2008 and the post-crisis high price stability period 2009-2014.  \label{fig:oil_weights}}
\end{figure}

The performance of the VAR model in Figure \ref{fig:oil_violin} also merits attention. It exhibits particularly strong performance during periods of high oil price volatility, such as 2003-2008 and 2020-2024. In contrast, its performance deteriorates markedly during the relatively stable period 2009-2014. This pattern is verified by the model weights shown in Figure \ref{fig:oil_weights}, which illustrates the weights of component models in the DTVW combination during 2003-2008 and 2009-2014. During the volatile period 2003-2008, the VAR model is assigned higher weights, whereas its weight decreases substantially during the more stable period 2009–2014. This phenomenon indicates that the DTVW can adaptively identify under-performing models, assigning them lower weights while prioritizing models with better predictive accuracy.

These findings suggest that the DTVW mechanism of dynamically adjusting weights to time-varying volatility and structural breaks is critical to modeling complex commodity markets such as oil. Its robustness in multiple metrics positions DTVW as a promising tool for policymakers and practitioners who require reliable probabilistic forecasts in volatile environments.

\subsection{GDP and Inflation}\label{sec-5.2}

In this subsection, we employ quarterly, seasonally adjusted U.S. data on personal consumption expenditures (PCE) inflation and real gross domestic product (GDP), both sourced from the Bureau of Economic Analysis (BEA). We jointly estimate GDP growth and inflation, with the dimension $L=2$ in Section \ref{sec-2.1}. Inflation is measured as the quarterly growth rate of the PCE deflator, calculated as $100(\log(\mathrm{PCE}_t)-\log(\mathrm{PCE}_{t-1}))$, covering the period from 1960:Q1 to 2009:Q4. The corresponding GDP growth rate is defined analogously as $100(\log(\mathrm{GDP}_t)-\log(\mathrm{GDP}_{t-1}))$, and spans the same sample period. In this subsection, we conduct out-of-sample, 1-step ahead forecasts for both GDP and PCE. This experiment serves two purposes: it highlights the model’s applicability in high-dimensional settings and demonstrates its capacity to generate accurate forecasts for macroeconomic variables.

We follow \citet{B:13} in using the same dataset and individual forecast methods, comprising six state-of-the-art time series models: a univariate autoregressive (AR) model, a bivariate vector autoregressive (VAR) model, and their Markov-switching variants (ARMS and VARMS). In addition, we include the time-varying parameter counterparts with stochastic volatility, namely TVPARSV and TVPVARSV. These models span a broad range of structural forms—linear, regime-switching, and time-varying—thereby ensuring substantial diversity in predictive behavior. Model parameters are estimated over the period 1960:Q1–1969:Q4. For each candidate model, we generate 1,000 independent simulations to approximate the predictive likelihood over the forecast evaluation period 1970:Q1–2009:Q4.

\begin{figure}[htbp]
\centering
\includegraphics[width=0.45\linewidth]{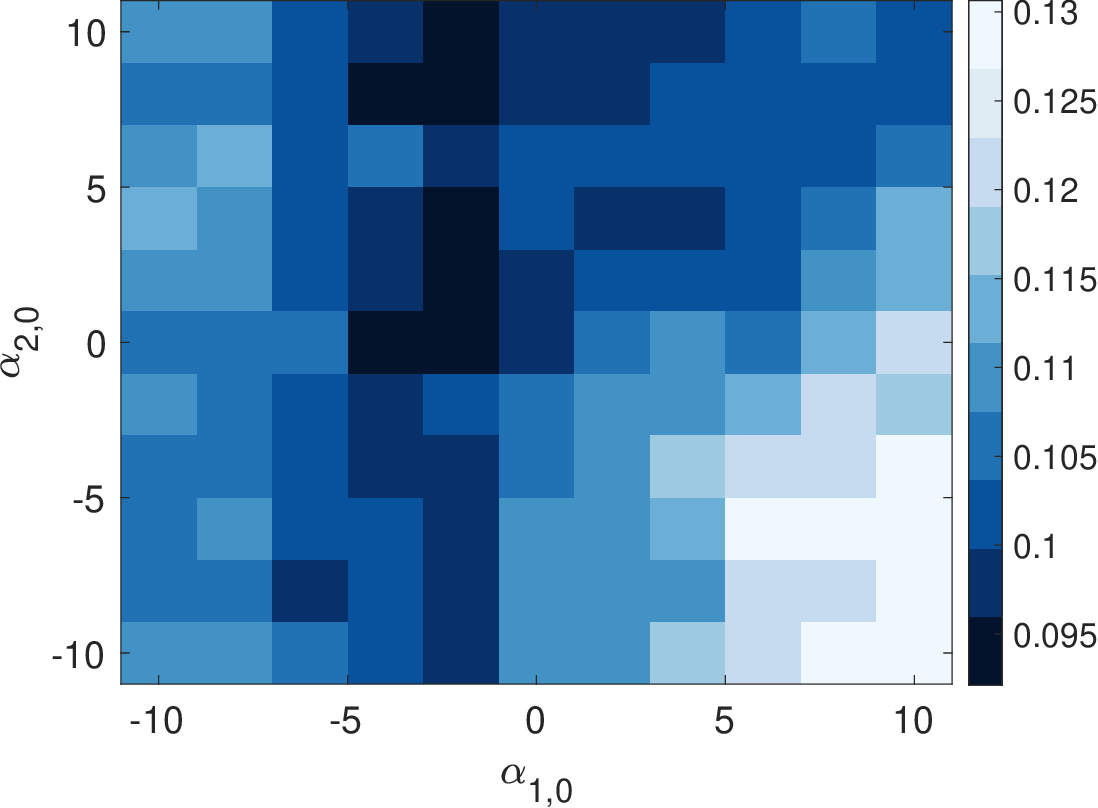}
\includegraphics[width=0.45\linewidth]{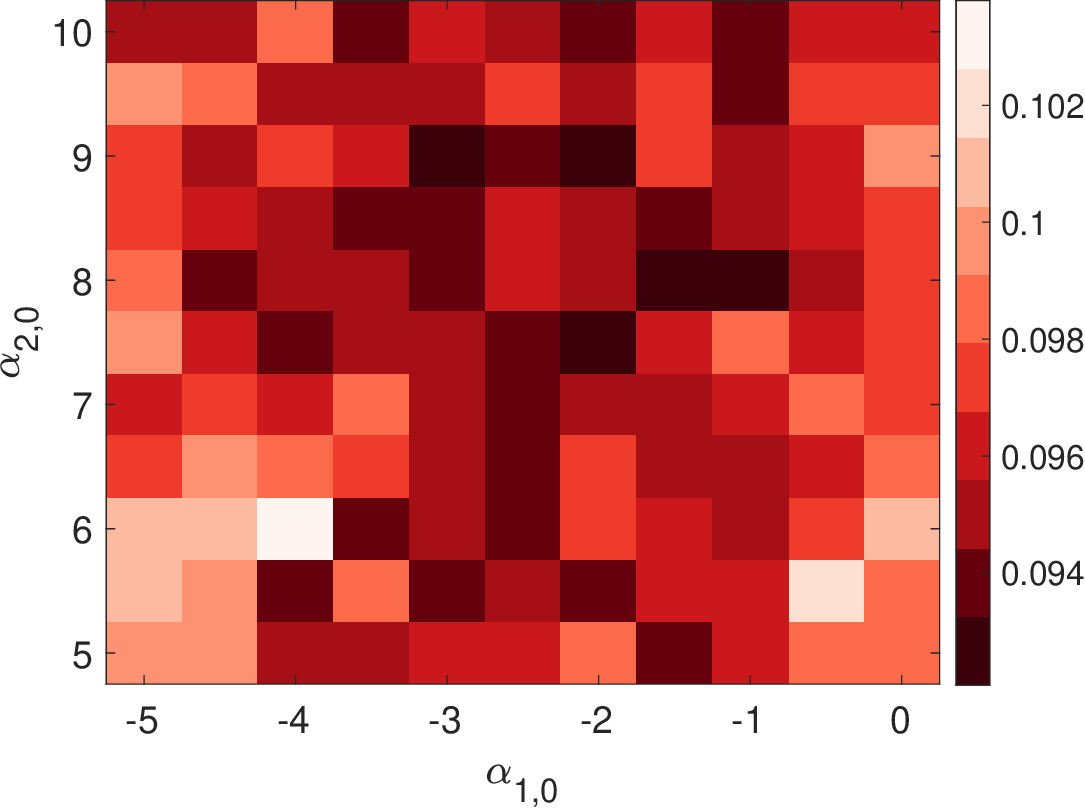}
\caption{Two-stage grid search for the initial values $(\alpha_{1,0},\alpha_{2,0})$ in the DTVW for the CRPS.}
\label{fig:grid_macro}
\end{figure}

We employ five combination methods, including Equal, BMA, BMA\_roll, TVW, and our DTVW. Equal initial weights are assigned to the six individual forecast models. The initial parameters $\alpha_{1,0}$ and $\alpha_{2,0}$ in DTVW are selected via the two-step grid search based on the CRPS performance for PCE as described in Section \ref{sec-4.2.2}. The heatmaps are shown in Figure \ref{fig:grid_macro}, suggesting that $(\alpha_{1,0},\alpha_{2,0})=(-2,9)$. 

\setlength{\tabcolsep}{3pt}
\begin{table}[H]
    \centering
    \small
    \caption{Forecast accuracy for the macro application}
    \resizebox{\textwidth}{!}{%
        \begin{minipage}{1.4\textwidth} 
            \centering
            \begin{tabular}{lcccccc@{\hspace{15pt}}ccccc}
                 \hline
        \textbf{PCE} &AR& ARMS& TVPARSV& VAR& VARMS & TVPVARSV & Equal &BMA & BMA\_roll & TVW& DTVW \\
        \hline
        RMSFE & 0.387& 0.387& 0.371& 0.388$^{**}$& 0.750& 0.384& 0.392&0.375& 0.363$^{*}$& 0.252$^{**}$& \textbf{0.227}$^{**}$  \\
        LS & 0.587 & 0.619 & 0.376 &   0.638 & 1.091$^{*}$ & 0.595 &0.800& 1.493$^{**}$ & 1.238$^{*}$ & -0.639$^{**}$ &  \textbf{-0.914 }$^{**}$ \\
        CRPS & 0.202& 0.199& 0.196& 0.203$^{**}$& 0.376& 0.201&0.204& 0.203& 0.196&0.106$^{**}$&\textbf{0.088}$^{**}$ \\
        \hline \hline
\textbf{GDP}  &AR& ARMS& TVPARSV& VAR& VARMS & TVPVARSV &Equal & BMA & BMA\_roll & TVW& DTVW \\
\hline
    
        RMSFE & 0.885 & 0.905&    0.852$^{**}$&  0.875$^{*}$&  0.976&   0.870& 0.865&0.857$^{*}$& 0.851$^{*}$&    0.628$^{**}$&  \textbf{ 0.619}$^{**}$ \\
        LS &1.320 &	1.405 &	1.185$^{**}$ &	1.377 &	1.362$^{*}$ &	1.225 &		3.183$^{**}$ &2.904$^{**}$ &1.764 & 0.794$^{**}$& \textbf{0.772}$^{**}$\\
        CRPS &  0.480& 0.472& 0.446$^{**}$& 0.468$^{*}$& 0.524$^{*}$& 0.453& 0.480 &0.474& 0.457$^{*}$&\textbf{0.300}$^{**}$&0.301$^{**}$ \\
        \hline 
        
            \end{tabular}
            \\[3pt]
            \raggedright Note: Bold numbers show the most accurate forecasts. $*$ indicates the result is better than the no-change model at $5\%$ confidence level using Diebold-Mariano test, and $**$ is at $1\%$ level.
        \end{minipage}
    }
    \label{tab:Macro_accuracy}
\end{table}

The results of the model comparison are summarized in Table \ref{tab:Macro_accuracy}, with DM tests conducted to assess predictive accuracy. Our DTVW consistently outperforms both individual forecast models and alternative combination schemes in PCE forecast, achieving the lowest RMSFE, LS, and CRPS values across all benchmarks. In GDP forecast, DTVW also yields the lowest RMSFE and LS. However, due to the two-dimensional estimation framework—where GDP and PCE are jointly estimated via particle filtering and thus mutually influential—the DTVW exhibits slightly inferior CRPS performance. This may be attributed to the initial parameter settings, which were optimized for PCE forecast CRPS, potentially compromising GDP prediction accuracy to prioritize improvements in PCE performance.

\begin{figure}[h!]
\begin{center}
\includegraphics[width=5in]{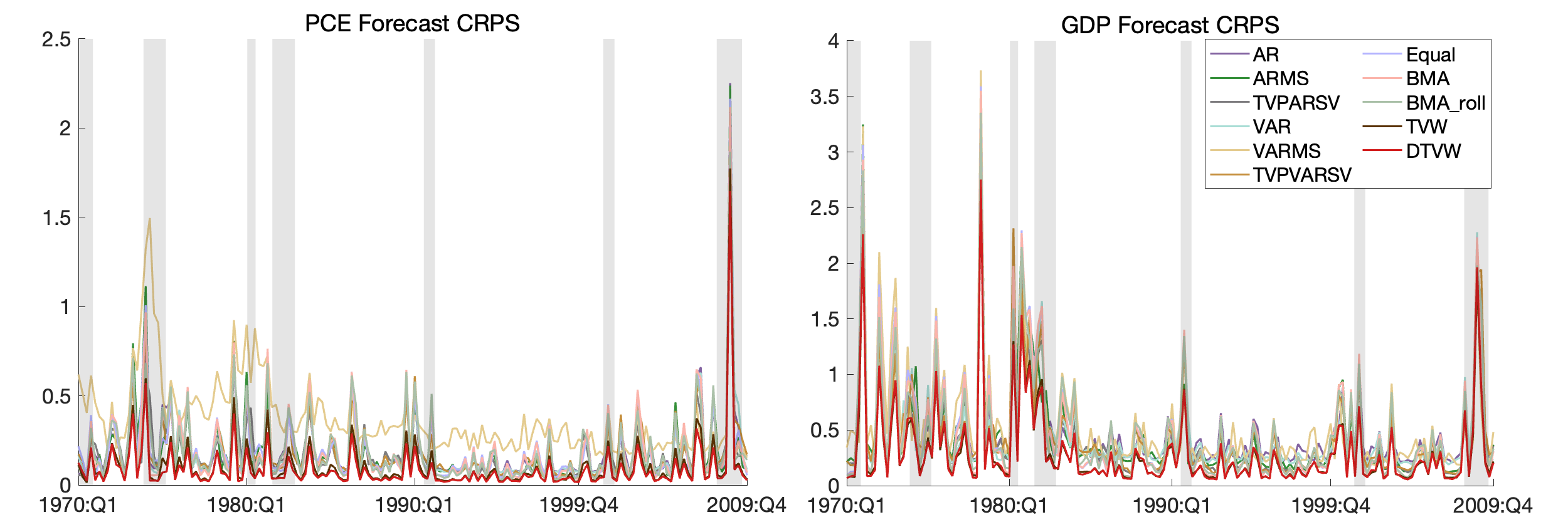}
\end{center}
\caption{The CRPS of PCE and GDP over the forecast evaluation period 1970:Q1-2009:Q4. Note: Gray areas indicate NBER bussiness cycle recessions dates.}
\label{fig:crps_Macro}
\end{figure}

Figure \ref{fig:crps_Macro} plots the CRPS trajectories of all forecast methods over the evaluation period, revealing two key patterns. First, the DTVW method consistently yields the lowest CRPS values at nearly every time point, underscoring its superior predictive calibration compared to competing approaches. For GDP, the CRPS results of DTVW and TVW are nearly identical, with only marginal differences. Second, the CRPS trajectories exhibit distinct temporal heterogeneity: forecast errors tend to be lower and more stable during periods of macroeconomic stability-such as the ``Great Moderation" (1982-2007)-yet rise substantially during turbulent episodes, including the 1970s stagflation, early-1980s recessions, and 2008 global financial crisis. Notably, DTVW’s enhanced performance reflects its capacity to leverage model diversity, generating more accurate and stable forecasts across varying economic regimes. These findings highlight the value of diversity-aware forecast combination strategies, particularly in strengthening robustness during volatile periods.

\begin{figure}[h!]
\begin{center}
\includegraphics[width=5in]{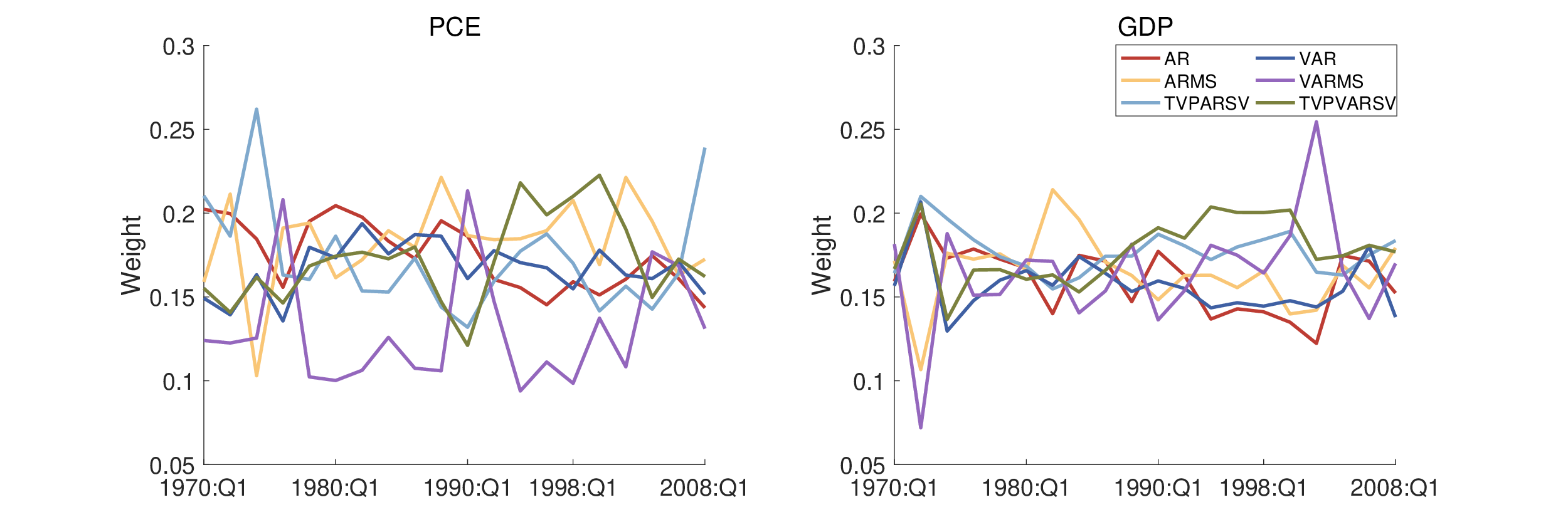}
\end{center}
\caption{Mean weights of individual models in the DTVW for PCE and GDP.
\label{figwei_macro}}
\end{figure}

Figure \ref{figwei_macro} illustrates the time-varying weights of individual models under the DTVW. For PCE forecast, the VARMS model is assigned relatively low and volatile weights, which aligns with its poorer forecast performance: as documented in Table \ref{tab:Macro_accuracy}, its CRPS value for PCE is approximately $0.38$, respectively—around 47\% higher than those of other models. In contrast, for GDP forecast, the TVPARSV and TVPVARSV models receive higher and more stable weights, consistent with their superior predictive accuracy. These results highlight that the weighting scheme effectively mirrors the relative forecast performance of individual models: it allocates greater weights to more accurate models and lower weights to less accurate ones. This is consistent with the findings in Section \ref{sec-4.2.2}, which demonstrate that DTVW can systematically identify and prioritize models with higher forecast accuracy.

\subsection{Investigation of the Feedback Mechanism in the DTVW}

Deeper insights into the forward-looking feedback mechanism of DTVW emerge from further observations on the estimated time-varying parameters $\alpha_{1,t}$ and $\alpha_{2,t}$ in the latent weight evolution Equation~\eqref{eqn-3.8}. These observations not only enhance understanding of why employing forecast diversity as a proxy for future information is both conceptually rigorous and empirically valid but also clarify how DTVW balances historical information against forward-looking forecast information across varying environments.

\setlength{\textfloatsep}{8pt}
\begin{figure}[h!]
\setlength{\abovecaptionskip}{4pt}
\centering

\begin{minipage}[t]{0.95\textwidth}
    \centering \footnotesize Simple complete model Equation~\eqref{eqn-4.1}
    \includegraphics[width=\linewidth,
    height=0.13\textheight]{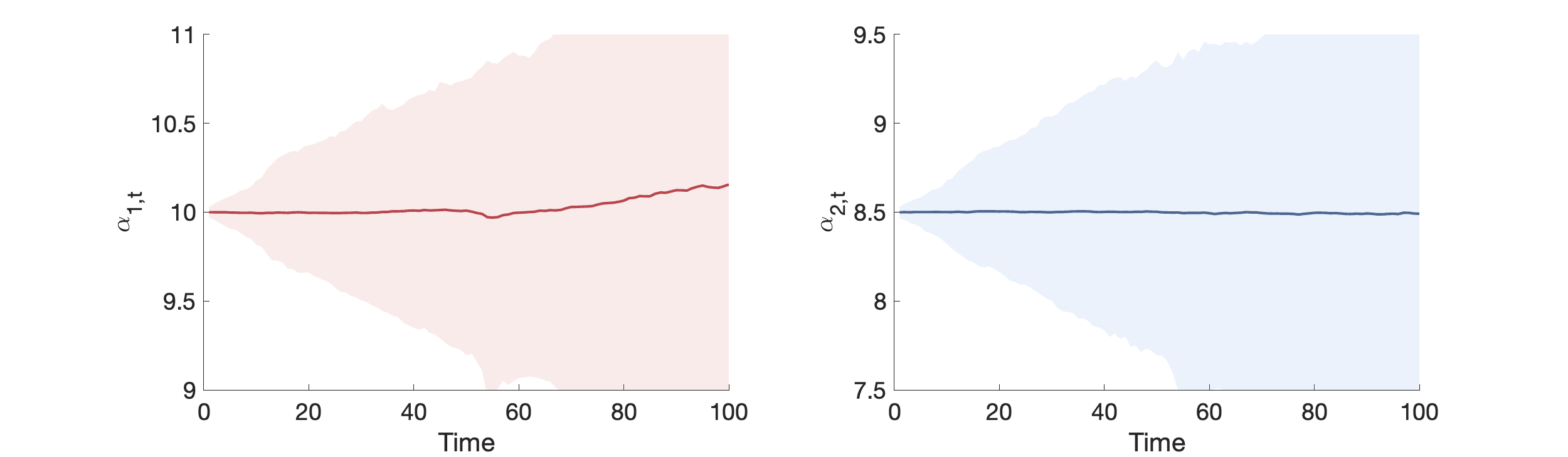}
    \small 
\end{minipage}
\vspace{4pt}

\begin{minipage}[t]{0.95\textwidth}
    \centering \footnotesize Complex incomplete model Equation~\eqref{eqn-4.2}
    \includegraphics[width=\linewidth,
    height=0.13\textheight]{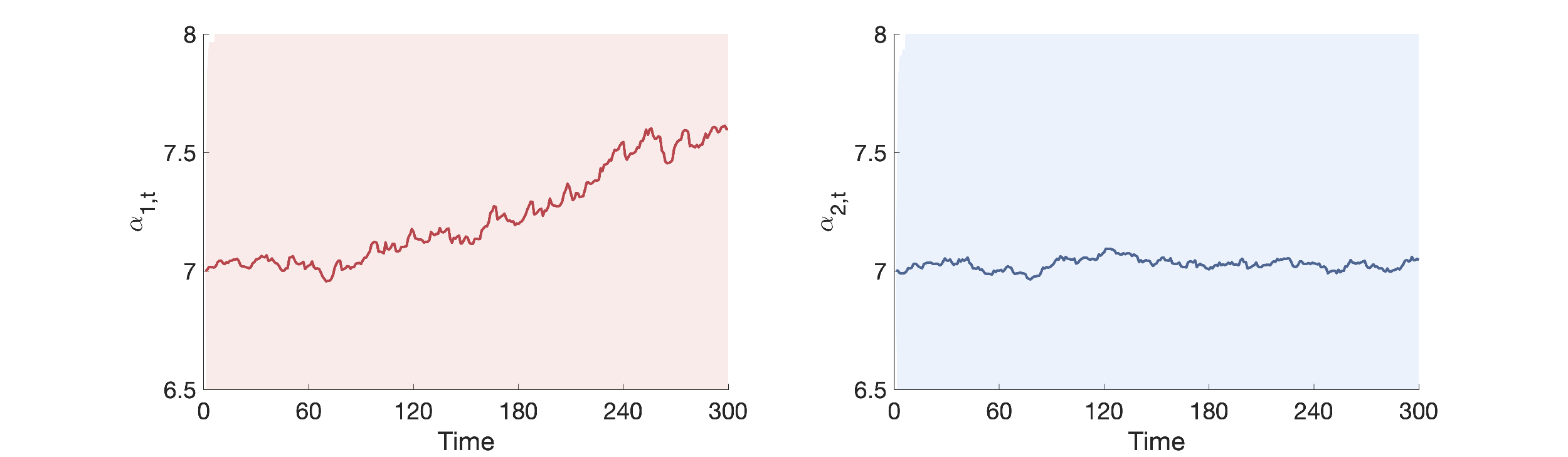}
    \small 
\end{minipage}
\vspace{4pt}

\small
\caption{Posterior means of time-varying parameters $(\alpha_{1,t},\alpha_{2,t})$ in the latent process Equation~\eqref{eqn-3.8}-\eqref{eqn-3.10} for two numerical simulations in Section \ref{sec-4.2}; 95\% confidence intervals are shaded.}
\label{fig:simucoefficient}
\end{figure}

Figure \ref{fig:simucoefficient} displays the time-varying estimated parameters $\alpha_{1,t}$ and $\alpha_{2,t}$ in the DTVW of the simulations in Section \ref{sec-4.2} with the same initialization selected there. In the simple complete model Equation~\eqref{eqn-4.1}, the parameters remain substantially stable over time, with 95\% confidence intervals that are relatively narrow compared to those of the complex incomplete model Equation~\eqref{eqn-4.2}. In contrast, both $\alpha_{1,t}$ and $\alpha_{2,t}$ in the complex incomplete model Equation~\eqref{eqn-4.2} exhibit more pronounced fluctuations, accompanied by considerably wider 95\% confidence intervals. These differences indicate that the simple complete model structure produces more precise and stable parameter estimates, reflecting lower posterior uncertainty. Conversely, the increased complexity and misspecification in the complex incomplete model introduce greater uncertainty into the latent process.

\setlength{\textfloatsep}{8pt}
\begin{figure}[h!]
\setlength{\abovecaptionskip}{4pt}
\centering
\begin{minipage}[t]{0.95\textwidth}
    \centering \footnotesize Oil Price
    \includegraphics[width=\linewidth,
    height=0.13\textheight]{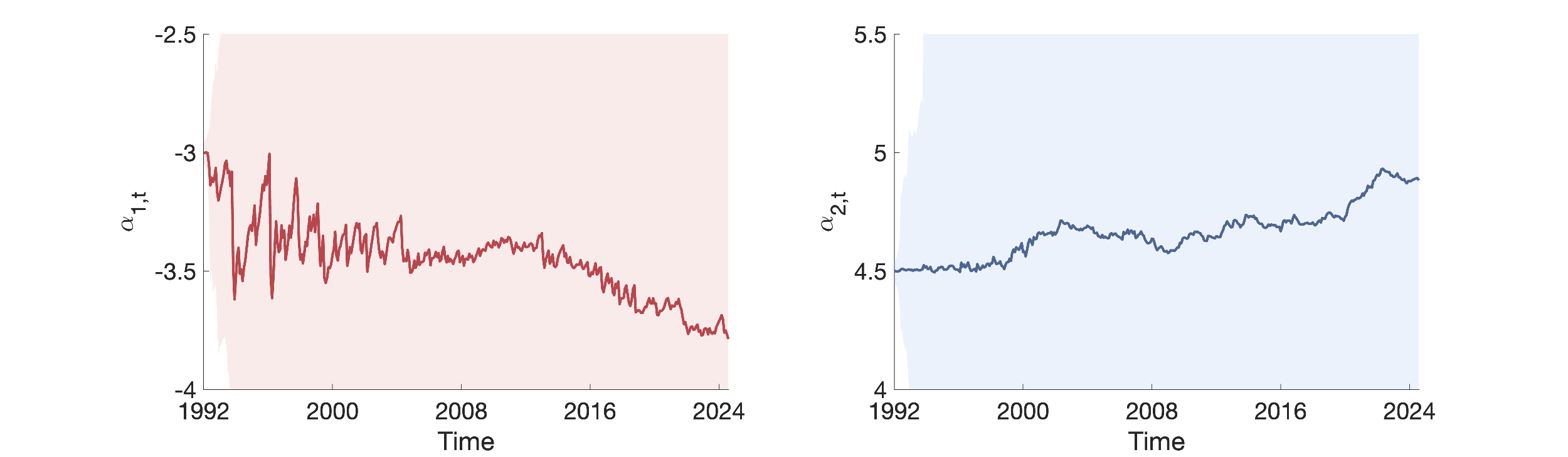}
    \small 
\end{minipage}
\vspace{4pt}

\begin{minipage}[t]{0.95\textwidth}
    \centering  \footnotesize PCE
    \includegraphics[width=\linewidth,
    height=0.13\textheight]{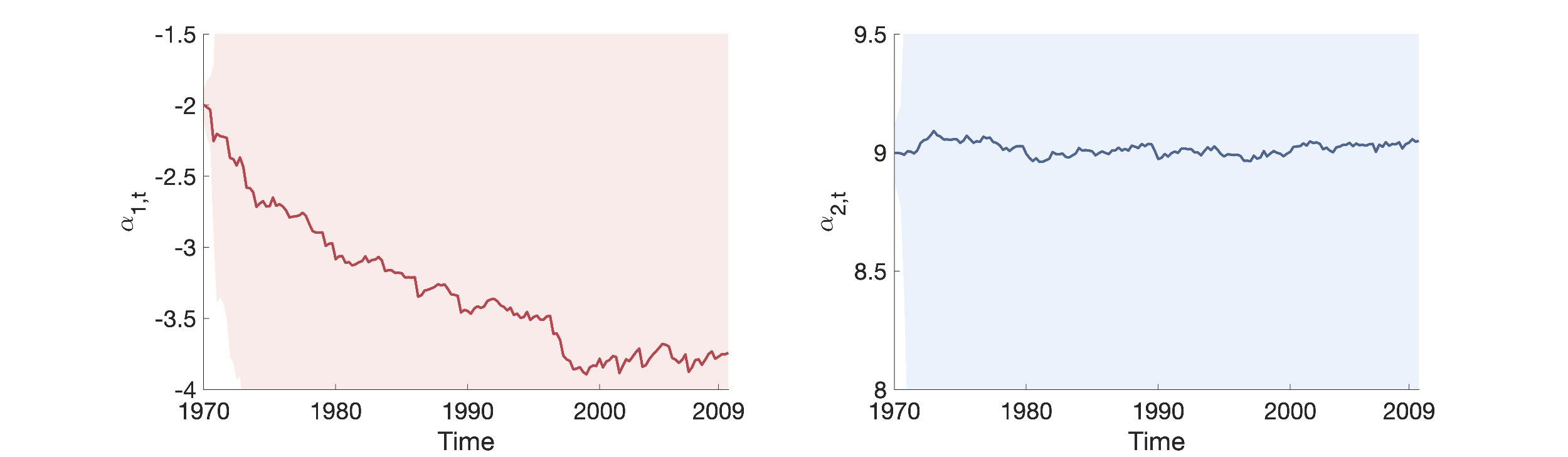}
    \small 
\end{minipage}
\vspace{4pt}

\begin{minipage}[t]{0.95\textwidth}
    \centering \footnotesize GDP
    \includegraphics[width=\linewidth,
    height=0.13\textheight]{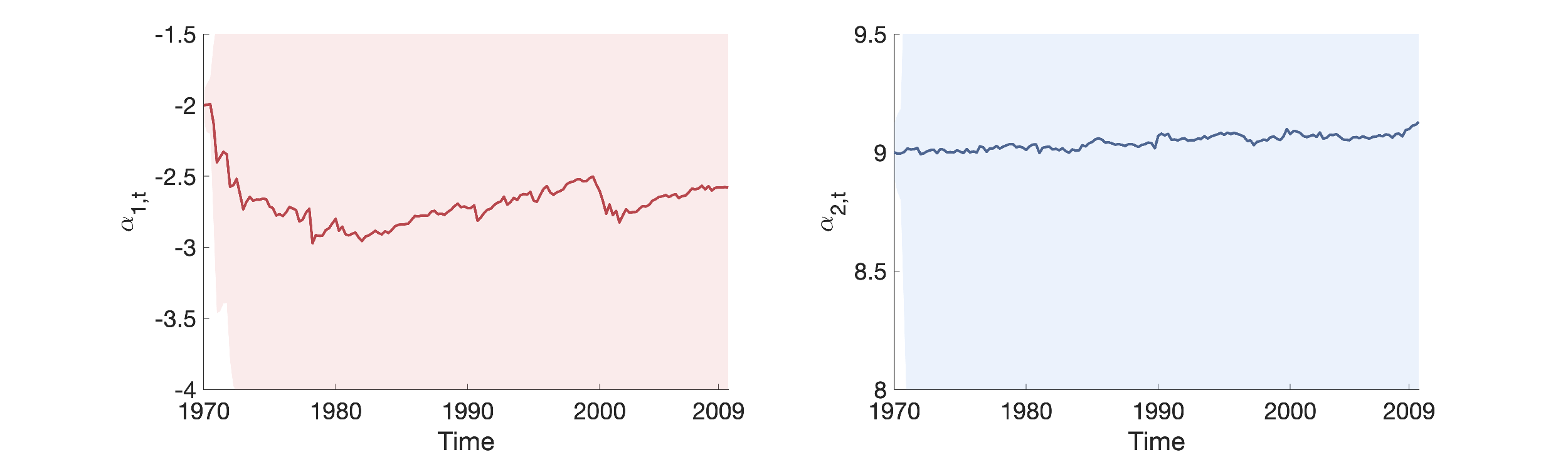}
    \small 
\end{minipage}

\vspace{4pt}
\small
\caption{Posterior means of time-varying parameters $(\alpha_{1,t},\alpha_{2,t})$ in the latent process Equation~\eqref{eqn-3.8}-\eqref{eqn-3.10} for two empirical applications in Section \ref{sec-5.1} and \ref{sec-5.2}; 95\% confidence intervals are shaded.}
\label{fig:empcoefficient}
\end{figure}

In empirical applications, patterns similar to those observed in the complex incomplete model Equation~\eqref{eqn-4.2} emerge across all three cases presented in Section \ref{sec-5.1} and \ref{sec-5.2}, as illustrated in Figure \ref{fig:empcoefficient}. The estimated parameters exhibit notable temporal fluctuations rather than remaining stable, with 95\% confidence intervals spanning considerably wide ranges. Notably, the parameters for oil prices display even more pronounced fluctuations, likely reflecting the nonstationary nature of real oil prices, which tend to undergo sharper changes compared to macroeconomic variables. Consistent with the simulation results in Figure \ref{fig:simucoefficient}, it is noteworthy that $\alpha_{2,t}$ remains positive across all empirical examples. However, $\alpha_{1,t}$ turns negative. This can be attributed to the fact that while the simulation design involves a complex incomplete model, the data are still generated from a well-specified process, lending credibility to the influence of historical information on $\alpha_{1,t}$. In contrast, empirical applications lack an exact data-generating process. Consequently, the model appears to distrust historical signals, assigning a negative value to $\alpha_{1,t}$ and exhibiting a downward trend over time.

\begin{figure}[h!]
\centering

\begin{subfigure}[t]{0.49\textwidth}
    \vspace{0pt}
    \centering \footnotesize Simple complete model Equation~\eqref{eqn-4.1}
    \includegraphics[width=\linewidth]{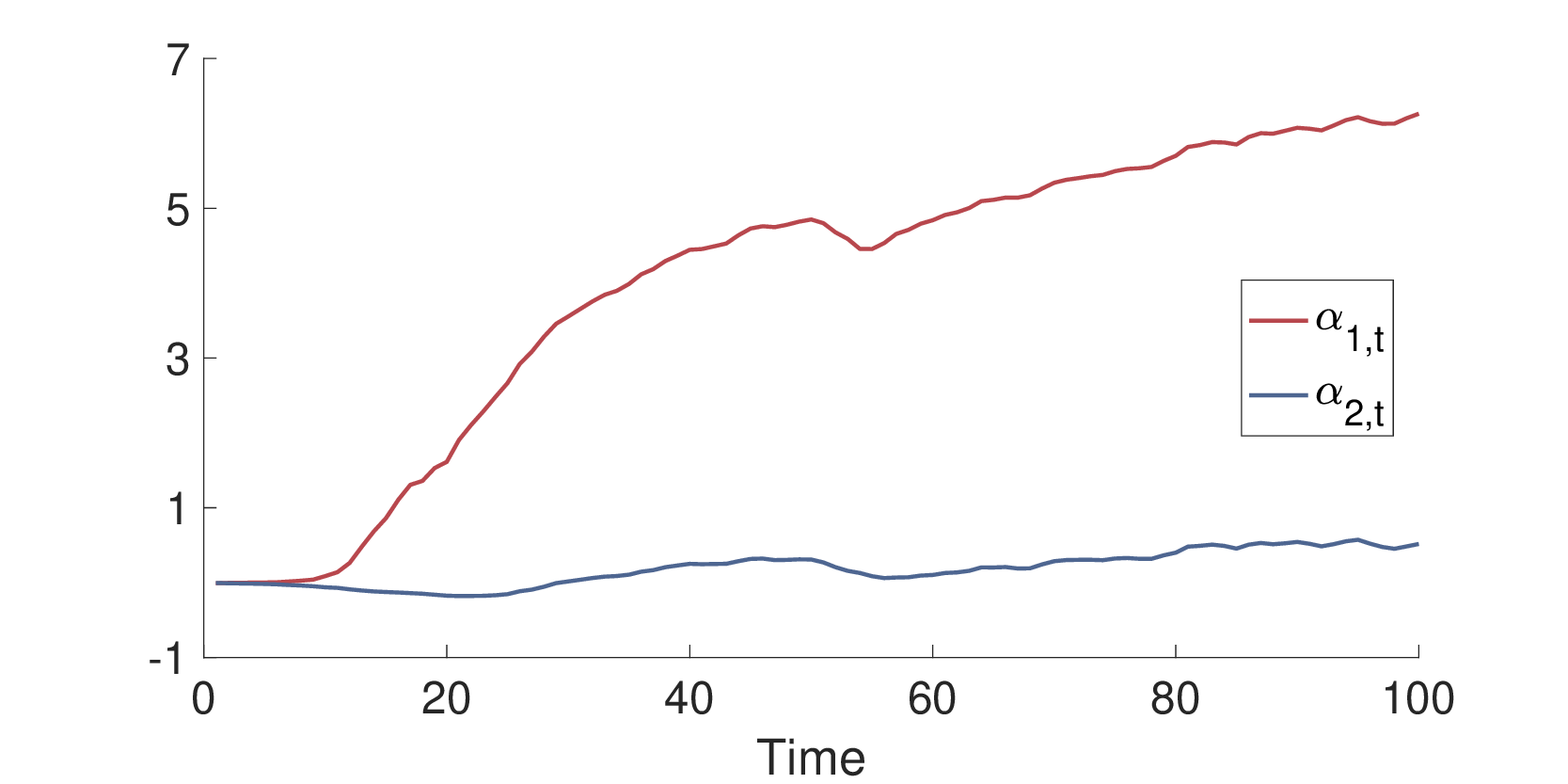}
    \label{fig:simu00}
\end{subfigure}
\hfill
\begin{subfigure}[t]{0.49\textwidth}
    \vspace{0pt}
    \centering \footnotesize Oil Price
    \includegraphics[width=\linewidth]{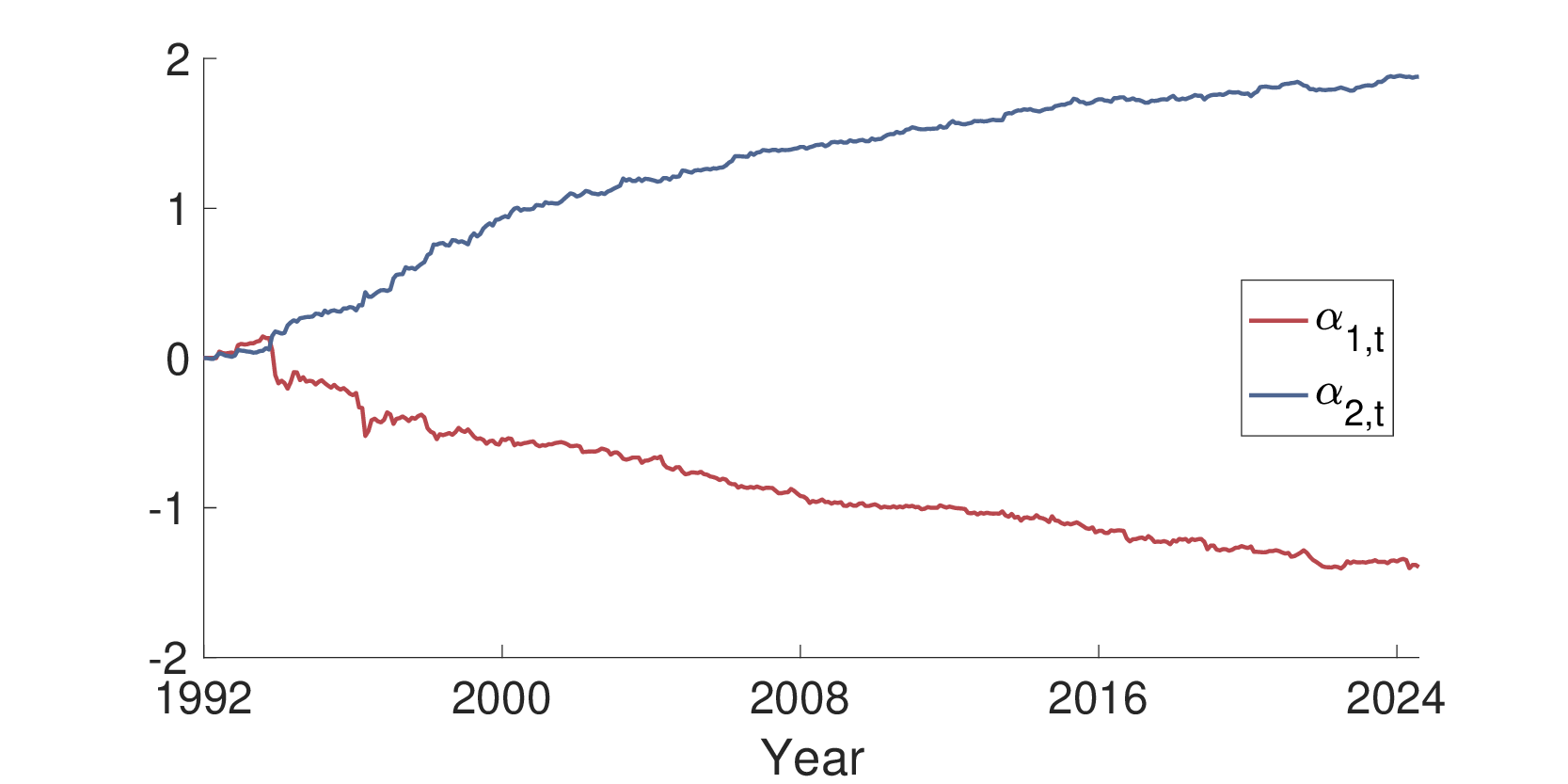}
    \label{fig:oil00}
\end{subfigure}
 \vspace{-20pt}
\caption{Time-varying parameters starting from zeros in the simple complete model Equation~\eqref{eqn-4.1} and oil price.}
\label{fig:5.300}
\end{figure}

To verify that the parameters are determined by the data itself rather than initial parameter values, we conducted an additional experiment. Specifically, instead of selecting initial values via grid search as in previous experiments, we set both $\alpha_{1,0}$ and $\alpha_{2,0}$ to zero and estimated them using the DTVW  applied to the simple complete model Equation~\eqref{eqn-4.1} and the oil price dataset. The results are presented in Figure \ref{fig:5.300}. For the simple complete model Equation~\eqref{eqn-4.1}, the parameter $\alpha_{2,t}$, associated with diversity, remains positive but close to zero over time, while $\alpha_{1,t}$, linked to historical observations, increases rapidly. In contrast, for the oil price application, $\alpha_{2,t}$ rises whereas $\alpha_{1,t}$ turns negative. This pattern confirms that in the DTVW the data itself automatically drives the model to prioritize either historical information or forecast diversity.

Whether in simulations or empirical studies, the parameter $\alpha_{2,t}$-which governs the role of diversity-consistently takes positive values. This indicates that models with greater diversity systematically receive higher weights in the DTVW. This finding aligns with the intuition that enhancing diversity enables forecast combinations to integrate a broader range of predictive information, thereby improving robustness against model misspecification and structural changes.

\section{Conclusion}\label{sec-6}

This paper introduces a novel forward-looking feedback framework, which leverages not only historical data but also anticipatory signals from future to inform the weight estimation process. What distinguishes this framework is its use of forward-looking information to form a predictive prior within a Bayesian structure: predictions extending beyond the current time point are used to refine the learning of weight dynamics. By incorporating these forward-looking components, the framework enables the model to adapt dynamically to anticipated structural changes.

The choice of a forward-looking signal is flexible and can be tailored to different forecast perspectives. In this study, we select the model diversity as the forward-looking signal and propose the DTVW approach. The main idea is to treat the model diversity, the differences among 1-step or multi-step ahead predictions from individual models, as a dynamic source of information. In doing so, the method effectively penalizes redundant or highly correlated models while reinforcing those that offer distinct and informative contributions to the overall forecast. In this approach, model weights evolve according to a latent process formulated as a regression, with all coefficients estimated sequentially from the data. This allows for a balanced integration of both types of information, resulting in more robust and reliable forecasts.

The proposed DTVW consistently delivers superior forecast performance in both simulation studies and real-world applications. In simulation experiments, we find that DTVW can successfully identify the true model when it is included in the candidate set; when the true model is absent, the method effectively downweights poorly performing models while favoring those that provide more accurate and complementary information.

In empirical applications, DTVW shows strong predictive ability across datasets with markedly different characteristics. The oil price dataset is characterized by pronounced volatility and clear structural breaks, making it a challenging forecast target. This experiment demonstrate that DTVW achieves the best performance among all competitors, including TVW and BMA, across all horizons, with its advantage in probabilistic forecasts becoming even more pronounced as the horizon lengthens. Consistent with the simulation findings,VAR received higher weights in the volatile early 2000s (for stronger performance) but lower weights in the stable period 2009-2014 (due to weaker accuracy). The GDP and PCE datasets are relatively stable and lend themselves to bi-variate modeling, providing a contrasting setting to the highly volatile oil price series. For both series, DTVW achieves competitive forecast performance. In GDP forecasting, the observed slight underperformance of CRPS arises because the initial parameter selection was predicated on minimizing the CRPS of PCE. Consequently, within the context of bivariate estimation, the predictive performance of GDP is liable to be impaired. In comparison, oil price forecasts outperform macroeconomic ones, probably because oil price forecasting is more complex, and DTVW works better for more intricate systems. The weight allocations remain reflective of model performance (e.g., downweighting underperforming VARMS in PCE, prioritizing stronger models like TVPARSV in GDP). These patterns mirror the simulation results, where the framework effectively allocates weights in accordance with model performance under different data conditions. By incorporating predictive priors that anticipate future changes, the method produces more reliable probabilistic forecasts, offering a flexible and robust tool for decision-making under uncertainty.

Future research could enhance the forward-looking feedback framework by incorporating richer forms of feedback, such as forecast disagreement across horizons or scoring rules. Extensions to accommodate macro-financial uncertainty indicators or external signals (e.g., volatility indices, sentiment scores) may further improve its responsiveness to structural changes. Such developments would broaden its applicability in real-time decision-making and complex policy environments.

\section*{Acknowledgements}

The authors gratefully acknowledge financial support from the National Natural Science Foundation of China (No. 12271019 and No. 72171011) and the National Key R\&D Program of China (2022YFA1005103).

\bibliographystyle{elsarticle-harv}
\bibliography{references}

\end{document}